 \documentclass[aps,pra,nofootinbib,showpacs]{revtex4}

 \usepackage{amsmath,amssymb,bm,euscript}
 \usepackage{graphicx}
 \usepackage[bf,small]{caption2}
 \usepackage{floatflt}

\def\WRM{Waves Random Media\ }

\def\JPCM{J. Phys.: Condens. Matter\ }
\def\RPP{Rep. Prog. Phys.\ }
\newcommand{\bra}[1]{\ensuremath{\bm{\langle}#1\bm{|}}}
\newcommand{\ket}[1]{\ensuremath{\bm{|}#1\bm{\rangle}}}

\newcommand{\sgn}{\ensuremath\mathrm{\,sgn\,}}

\begin{document}

\title{Metal-insulator transition in a two-dimensional electron system:\\
       the orbital effect of in-plane magnetic field}

 \author{Yu.\,V. Tarasov}
 \email{yutarasov@ire.kharkov.ua}
 \affiliation{Institute for Radiophysics and Electronics,
              National Academy of Sciences of Ukraine,
              12 Acad. Proskura St., Kharkov 61085, Ukraine}

\date{\today}

\begin{abstract}
 The conductance of an open quench-disordered two-dimensional (2D) electron
 system subject to an in-plane magnetic field is calculated within the
 framework of conventional Fermi liquid theory applied to actually a
 three-dimensional system of spinless electrons confined to a highly
 anisotropic (planar) near-surface potential well. Using the calculation
 method suggested in this paper, the~magnetic field piercing a finite range
 of infinitely long system of carriers is treated as introducing the
 additional highly non-local scatterer which separates the circuit thus
 modelled into three parts --- the system as such and two perfect leads. The
 transverse quantization spectrum of the inner part of the electron waveguide
 thus constructed can be effectively tuned by means of the magnetic field,
 even though the least transverse dimension of the waveguide is small
 compared to the magnetic length. The initially finite (metallic) value of
 the conductance, which is attributed to the existence of extended modes of
 the transverse quantization, decreases rapidly as the magnetic field grows.
 This decrease is due to the mode number reduction effect produced by the
 magnetic field. The closing of the last current-carrying mode, which is
 slightly sensitive to the disorder level, is suggested as the origin of the
 magnetic-field-driven metal-to-insulator transition widely observed in 2D
 systems.
\end{abstract}


\maketitle %
\section{Introduction}

The conduction properties of low-dimensional electron and hole systems with
the disorder of different origin have long been the subject of active
research. Investigations into such objects of mesoscopic size have currently
become particularly intensive in view of their applied importance
(semiconductor heterostructures, quantum dot devices, etc.), on the one hand,
and due to the intriguing uncommonness of the obtained results, on the other.
The most puzzling phenomenon which has not as yet been understood in full
measure is the transition of 2D electron systems from conducting to
insulating state (the extensive bibliography on the subject can be found in
Refs.~\cite{bib:AbKrSar01,bib:KrSar04}). Observation of this phenomenon
evidently contradicts the common view stemming from the well-known scaling
theory of localization~\cite{bib:AALR79}.

A diversity of ideas have been put forward to explain the ``anomalous''
metallic behaviour of 2D systems at extremely low temperatures. Among
physical mechanisms assumed to be responsible for such behaviour the most
appropriate is the Coulomb ($e$-$e$) interaction of carriers. The assumption
is mainly based on the fact that the dimensionless parameter $r_s$ (the ratio
of interaction energy to Fermi energy of electrons) which characterizes this
interaction amounts to the value of $\gtrsim 10$ in planar systems of Si
MOSFET-type as well as in GaAs/AlGaAs heterostructures, which would suffice
to raise serious doubts about the adequacy of Fermi-liquid approach with
regard to such systems.

Note that in condensed matter physics an important paradigm is the
quasi-particle notion. A~number of phenomena can be understood in terms of
weakly interacting quasi-particles, although Coulomb interaction between the
electrons is rather strong. The Fermi liquid description of quasi-particles
has for a long time been considered to be efficient in low dimensions in the
presence of disorder \cite{bib:LR85,bib:CKL87} until recent experiments have
challenged this view \cite{bib:KKFPdI94}. Serious problems have also arisen
when interpreting numerous experimental data on decoherence time saturation
at low temperature, reviewed in Ref.~\cite{bib:LB02}, as well as explaning
unexpectedly large persistent currents in normal metals~\cite{bib:Moh99}.

At this stage, the lack of a comprehensive theory for strongly correlated
electrons does not allow for certain conclusions to be made about the role of
$e$-$e$ interaction in the detected phenomena. Since in different existing
theories this interaction is evaluated quite differently, viz. from promoting
localization \cite{bib:AAL80,bib:TC89} to preventing its appearance
\cite{bib:F84,bib:BK94,bib:CCL98}, it would be quite reasonable at first to
describe the observed effects in frames of one-particle approach and then to
regard Coulomb interaction as the added dephasing factor. Only should one
fail in obtaining the proper result on this canonical way may it be rational
to think of Coulomb correlations as a prevalent cause of forming
the~continuous component in the electron spectrum of 2D systems.

Previously \cite{bib:Tar00}, the one-particle theory capable of explaining
the conducting ground state of weakly disordered finite-size 2D systems was
developed in terms of electron states pertinent to open quantum wells of
waveguide configuration. According to this theory, the conductance must
exhibit non-exponential coordinate dependence at any length of the system
provided that there exists more than one extended mode of transverse
quantization (so called waveguide mode), commonly referred to as the open
conducting channel~\cite{bib:Been97}. In Ref.~\cite{bib:Tar00} it was proven
that scattering by static random potential, even though it is of elastic
nature from the viewpoint of one-electron dynamics, may well result in
dephasing the initially coherent extended \emph{collective} modes. This leads
to the reduction in the primordially finite-value ballistic conductance
rather than it grows from the ``localized'' exponentially small value. The
primordial conductance is quantized in steps, each being equal to the
conductance quantum $G_0=e^2/\pi\hbar$, regardless of the length of the
bounded 2D system of carriers. Inclusion of imperfections allowing for
scattering between different extended waveguide modes results in the
conductance decrease from ballistic to diffusive value. The expression for
the diffusive conductance tends asymptotically to the standard Drude form if
the quantum waveguide is sufficiently wide. The above scenario of the
conducting state of a real 2D electron system is evidently opposite to the
one stemming from well-known scaling theory, in which the metallic behaviour
of the conductance can only result from some kind of dephasing of the
electron states initially localized by disorder.

Using the approach taken in Ref.~\cite{bib:Tar00}, in Ref.~\cite{bib:Tar03}
one-particle theory of the metal-insulator transition (MIT) in 2D systems was
developed. This theory provides a deeper insight into the peculiarities of
the effect observed experimentally. According to \cite{bib:Tar03}, (i) MIT
detected in Si-MOSFETs as well as in structures of GaAs/AlGaAs-type should be
regarded as a true quantum phase transition which takes place both in
disordered and perfect systems; (ii)~straight from the metallic side of the
transition the conductance reaches the value equal, by the order of
magnitude, to the standard conductance quantum (theoretically, in the limit
of an ideally perfect system, the conductance jumps exactly by $G_0$ as the
last conducting channel is closed).

The results \cite{bib:Tar00,bib:Tar03} were obtained in the Fermi-liquid
description of the electron system, without taking explicitly into account
Coulomb interaction of carriers as well as spin effects. Meanwhile, the very
fact that the \emph{extended} waveguide modes were used as the initial
quasi-particles (which have unlimited spatial extent, in contrast to
Anderson-localized states) permits hoping that the results of
Refs.~\cite{bib:Tar00,bib:Tar03} will hold true, at least qualitatively, even
on condition that Coulomb correlations are also taken into account.

Although the model of 2D MIT suggested in Ref.~\cite{bib:Tar03} correlate
well with the main experimental facts, it does not seem to be rather
convincing to make far-reaching conclusions. The problem of great concern,
which has not been touched upon in \cite{bib:Tar03} and has not been so far
explained theoretically is the unusual nature of the strong localizing effect
produced by the relatively weak in-plane magnetic field. Normally, the
magnetic field is known to suppress the conductance of bulk metallic samples
due to coupling to the orbital motion of electrons. However, if the field is
applied parallel to a confined system whose actual thickness is small as
compared to the magnetic length, it is natural to expect the orbital effect
to be highly suppressed. Only corrections from $e$-$e$ interaction of
carriers are retained in the case of spinless electrons~\cite{bib:LR85}.

With this prevailing point of view, it came as a surprise when dramatic
suppression of conductivity was observed in Si-MOSFETs subjected to the
in-plane magnetic field~\cite{bib:DKSK92}. The effect of this field was found
to be so significant that it causes the zero-field 2D metal to become an
insulator \cite{bib:SKSP97,bib:MZVSK01,bib:SKK01,bib:GMRPW02}. Inasmuch as
magnetic fields appropriate for such a transition are strong enough to
completely polarize electron spins, this prompted many of the researchers
(see, e.~g., Refs.~\cite{bib:SHPLRRSG98,bib:OHKY99,bib:Metal99}) to assert
categorically that it is exactly the spin polarization that should be
regarded as a physical origin of 2D MIT in a parallel magnetic field. This
point of view is also supported by the lack of a comprehensive theory of the
electron transport in real quasi-two-dimensional (Q2D) systems, i.~e. planar
systems of finite, though rather small, thickness. Although in some papers
(e.~g., \cite{bib:DSHw00,bib:MFA02}) the orbital effect of an in-plane
magnetic field was, in a way, analyzed as well, the results are still not
rated as fully convincing. This is probably concerned, on the one hand, with
the fact that the relatively simple model proposed in \cite{bib:DSHw00} is
incapable of explaining a quite abrupt transition over the magnetic field
between metallic and dielectric regimes. On the other hand, the results
obtained in Ref.~\cite{bib:MFA02} only exhibit the trend in behaviour of the
Q2D system conductance with a growth in the magnetic field, since the applied
calculation method enables one to analyze only weak-localization
\emph{corrections} in the metallic phase, thus being incapable of capturing
the~coarse effect such as MIT.

In this paper, the orbital effect of in-plane magnetic field upon open
quasi-2D electron systems of waveguide configuration is investigated using
the theoretic-field method previously employed in
Refs.~\cite{bib:Tar00,bib:Tar03}. The advantage of the approach is that it
enables one to take into account in a similar fashion all the one-electron
states, without distinguishing them by the kind of trajectories, which in Q2D
case can be either of sliding type (Levi flights) or those frequently
scattered at side boundaries of the system. Although the method is widely
applicable, here we will examine thoroughly the specific case of weak
electron-impurity scattering and weak magnetic fields (the appropriate
criteria are listed below).

The notion of weak magnetic field implies the certain relationship between
geometric parameters of the electron waveguide and the classical cyclotron
radius of electron trajectories. At the same time, in general it does not
imply low mixing of the electron states. The magnetic-field-induced mode
entanglement may be rather strong, but it does not give rise to noticeable
dephasing of the transverse-quantized states unless there exists some random
short-correlated potential as well. The most important effect of the magnetic
field is that it influences significantly the mode content of the electron
spectrum, thereby resulting in noticeable change in the number of
current-carrying modes and in spectral width of their energy levels. In
Ref.~\cite{bib:Tar04prep}, it was proven that the number of conducting
channels becomes smaller when the magnetic field increases, being
simultaneously accompanied by the gain in coherence of the electron states.
It is just the reduction in the number of extended modes that can be viewed
as the physical origin of the transition of Q2D electron system from
conducting to insulating state.

To justify the sensitivity of the mode state spectrum as the major cause of
the observed MIT it is necessary to trace the spectrum peculiarities in terms
of the observable quantity, say, the conductance. To accomplish this task, it
would be inconvenient to use the Landauer approach, since it does not look
into the coherence properties of the \emph{internal} spectrum of the system
in question, operating it as a scattering object integrally. Therefore, in
this study we apply the linear response theory \cite{bib:Kubo57} which
permits us to calculate the conductance on a microscopic level.

\section{The model}

\begin{figure}[h]
\centering\vspace{.5\baselineskip}
\scalebox{.7}[.7]{\includegraphics{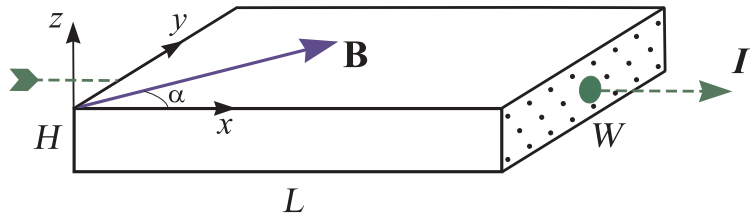}}
\captionstyle{hang}\caption{Configuration of the electron waveguide and the
magnetic field.\hfill\label{fig1}}
\end{figure}

Two-dimensional electron and hole systems in use, in view of their open
property in the direction of current and the resemblance of master equation
to that of the classical wave theory, can be simulated as planar quantum
waveguides whose transverse design is governed by the lateral confinement
potentials. Although near-surface potential wells in Si MOSFETs and
GaAs/AlGaAs heterostructures are close in shape to triangular or parabolic
form \cite{bib:BK94,bib:SarKrav99}, this fact is of minor importance for its
principal application which is to restrict electron transport in the
direction normal to heterophase areas, thus resulting in \emph{transverse}
quantization of the electron spectrum. With this consideration in mind, we
assume, to simplify calculations, the Q2D system of carriers to have the form
of three-dimensional planar ``electron waveguide'' of rectangular
cross-section (see Fig.~\ref{fig1}), which occupies the coordinate region
\begin{align}
 & x\in(-L/2,L/2)\ ,\notag\\
 & y\in[-W/2,W/2]\ ,\label{geom}\\
 & z\in[-H/2,H/2] \ .\notag
\end{align}
The length $L$, the width $W$, and the height $H$ of the waveguide will be
regarded as arbitrary, within the restrictions imposed below.

In practice the change of a quantum waveguide thickness implies alteration of
the width of the near-surface potential well and, as a consequence, of sheet
density of the carriers. Inasmuch as this density is known to follow the
variation in depletion voltage under the simple law
\cite{bib:S96,bib:WEPHAFRPJ89}, the results obtained below as a function of a
waveguide thickness can be easily related to the experiment.

We will examine the $T=0$ magnetoresistance of a Q2D electron system by
expressing the dimensionless (in units of $e^2/\pi\hbar$) conductance
$g_{xx}$ in terms of one-particle propagators. Assuming the system of units
with $\hbar=2m=1$ ($m$~is the electron effective mass), the static
conductance is given by
\begin{align}\notag
  g_{xx}(L,\mathbf{B}) = \frac{2}{ L^2}
  \iint\textit{d}\mathbf{r}\textit{d}\mathbf{r}'\,
  &\left[\frac{\partial}{\partial\,x}-\frac{ie}{\hbar c}A_x(\mathbf{r})\right]
  \left[G_{\varepsilon_F}^a(\mathbf{r},\mathbf{r}')-
  G_{\varepsilon_F}^r(\mathbf{r},\mathbf{r}')\right]\\
  \label{Kubo_cond}
   \times &\left[\frac{\partial}{\partial\,x'}-\frac{ie}{\hbar c}A_x(\mathbf{r}')\right]
  \left[G_{\varepsilon_F}^a(\mathbf{r}',\mathbf{r})-
  G_{\varepsilon_F}^r(\mathbf{r}',\mathbf{r})\right]\ ,
\end{align}
where $G_{\varepsilon_F}^{r(a)}(\mathbf{r},\mathbf{r}')$ is the retarded
(advanced) Green function of the electrons of Fermi energy~$\varepsilon_F$,
$A_i(\mathbf{r})$ is the $i$-th component of the external vector potential
$\mathbf{A}(\mathbf{r})$. Integration in \eqref{Kubo_cond} is performed over
the region \eqref{geom} occupied by the quantum waveguide, spin degeneracy is
taken into account by the factor of 2. Note that throughout this paper the
Fermi energy (or the chemical potential, as the situation requires) will be
considered to have a constant value, independently of the confinement
potential. This is undoubtedly true on the metallic side of the MIT discussed
below. Moreover, on just-dielectric side of the transition this assertion is
also valid since the ``expanded'' electron system, which includes the
attached leads, in the static limit is on a~homogeneous equilibrium state
provided that its segment of interest is open, at least, partially.

Within the model of isotropic Fermi liquid, the retarded Green function of
electrons subjected to a static magnetic field obeys the
equation (all indices are omitted for brevity) 
\begin{equation}\label{master_eq}
  \left[\left(\nabla-\frac{2\pi i}{\Phi_0}\mathbf{A}(\mathbf{r})\right)^2+
  k_F^2+i0-V(\mathbf{r})\right]G(\mathbf{r},\mathbf{r}')=
  \delta(\mathbf{r}-\mathbf{r}')\ ,
\end{equation}
where $\Phi_0=hc/e$ is the magnetic flux quantum, $k_F^2\equiv\varepsilon_F$,
$V(\mathbf{r})$ is the random potential due to impurities or the roughness of
the confining well boundaries. We will specify this potential by zero mean
value, $\big<V(\mathbf{r})\big>=\nolinebreak 0$, and the binary correlation
function
$\big<V(\mathbf{r})V(\mathbf{r}')\big>=\mathcal{QW}(\mathbf{r}-\mathbf{r}')$.
The angular brackets are used for configurational averaging,
$\mathcal{W}(\mathbf{r})$ is the function normalized to unity and falling off
its maximal value at $\mathbf{r}=0$ over the characteristic length $r_c$ (the
correlation radius).

Equation \eqref{master_eq} must be supplemented with the
appropriate boundary conditions (BC). We will regard the electrons
to be confined by infinitely high potential walls at side
boundaries of the region \eqref{geom} and specify this fact by the
Dirichlet conditions as follows,
\begin{equation}\label{Dirichlet}
  G(\mathbf{r},\mathbf{r}')\bigg|_{\genfrac{}{}{0pt}{}{y=\pm W/2}{z=\pm
  H/2}}=0\ .
\end{equation}
As far as open ends of the system are concerned, the BC problem is resolved
somewhat less trivially. The mere fact that the system is open, even
partially, implies non-Hermicity of the operator \eqref{master_eq} in the
domain \eqref{geom}. This may cause some vagueness regarding the
applicability of the formula \eqref{Kubo_cond}, whose derivation relies
essentially on Hermitian property of the Hamilton operator. In the case of a
finite-length system, the hermitizing BC at $x=\pm L/2$ would in fact
correspond to its closeness (or periodicity) in $x$-direction, which does not
conform with the requirement of current overflow between \emph{independent}
reservoirs.

In this study, in view of the chosen Green function formalism, to specify the
openness of the quantum system we employ the method based on the analogy
between the problem \eqref{master_eq} and that of the monochromatic point
source radiation in a classical waveguide. When solving the latter problem,
Sommerfeld radiation conditions are normally used \cite{bib:BF78,bib:Vlad67},
which imply for the source positioned at some finite coordinate the existence
of solely outgoing waves at infinity. In order to adapt these conditions to
the system under consideration it is necessary to prolong the disordered and
magnetic-field-subjected segment \eqref{geom} of the electron waveguide with
semi-infinite ideal leads, in which the electron waves generated at some
point inside the segment could propagate freely to infinity, not being
subjected to any kind of backscattering. Then, joining of the solutions to
Eq.~\eqref{master_eq} at interfaces $x=\pm L/2$ within the electron waveguide
results in the complete solution corresponding to the infinite open system,
thereby giving rise to the correct BC at the ends of the segment of interest.
This somewhat troublesome procedure will be done in the Appendix, where the
trial Green functions, which serve as a basis for the exact solution to
equation \eqref{master_eq}, are obtained.

Apart from the openness BC, one more problem is to be resolved before we
proceed to the conductance calculation. Specifically, this is the choice of
the vector potential gage corresponding to the in-plane magnetic field
$\mathbf{B}=(B_x,B_y,0)$. Here, the main motivation is to avoid quadratic
confinement in $x$-direction. The lack of such a confinement enables one to
employ the transverse eigenfunction basis consisting of waveguide modes, both
extended and evanescent, instead of normally used isotropic plane-wave basis.
The choice of the latter basis would make the BC at side boundaries not easy
to be satisfied, which would result in a substantially complicated solution.
We prefer the gage
\begin{equation}\label{gage}
  \mathbf{A}(\mathbf{r})=(B_yz,-B_xz,0)\ ,
\end{equation}
which yields equation \eqref{master_eq} of the form
\begin{equation}\label{main_eq}
  \Bigg[\nabla^2+k_F^2+i0-V(\mathbf{r})-\frac{4\pi i}{\Phi_0}\left(
  B_yz\frac{\partial}{\partial x}-B_xz\frac{\partial}{\partial y}\right)-
  \left(\frac{2\pi}{\Phi_0}\right)^2
  \mathbf{B}^2z^2\Bigg]G(\mathbf{r},\mathbf{r}')=
  \delta(\mathbf{r}-\mathbf{r}')\ .
\end{equation}
Eq.~\eqref{main_eq} is suitable for reducing the initially three-dimensional
stochastic problem to a set of more readily solved one-dimensional problems.
Such a reduction is particularly helpful in view of the openness BC, which is
most easily formulated in one spatial dimension.

\section{Reduction to one-dimensional dynamic problems}

Clearly, the initially three-dimensional dynamic problem might be reduced to
one-dimensional problems if one succeeded, yet hypothetically, in finding the
actual set of decoupled transverse eigenmodes of the \emph{random}
Hamiltonian in Eq.~\eqref{main_eq}. As a matter of fact, one can do so,
though indirectly, in terms of standard Fourier transformation of this
equation over transverse vector coordinate $\mathbf{r}_{\perp}=(y,z)$. In our
confinement model, the appropriate eigenfunctions are
\begin{equation}\label{transv_set}
  \ket{\mathbf{r}_{\perp};\bm{\mu}}=\frac{2}{\sqrt{WH}}
  \sin\left[\left(\frac{y}{W}+\frac{1}{2}\right)\pi n\right]
  \sin\left[\left(\frac{z}{H}+\frac{1}{2}\right)\pi m\right] \ ,
\end{equation}
with $\bm{\mu}=(n,m)$ being the vector mode index conjugate to
$\mathbf{r}_{\perp}$ ($n,m\in\mathbb{N}$). With functions \eqref{transv_set},
equation \eqref{main_eq} is transformed to the set of one-coordinate, yet
strongly coupled, equations for the mode Fourier components of the function
$G(\mathbf{r},\mathbf{r}')$, viz.
\begin{equation}\label{mode_eqn}
  \left[\frac{\partial^2}{\partial x^2}+
  k^2_{\bm{\mu}}+i0-
  {\mathcal V}_{\bm{\mu}}(x)\right]G_{\bm{\mu}\bm{\mu}'}(x,x')
  -\sum_{\bm{\nu}\neq\bm{\mu}}\hat{{\mathcal U}}_{\bm{\mu}\bm{\nu}}(x)
  G_{\bm{\nu}\bm{\mu}'}(x,x')=
  \delta_{\bm{\mu}\bm{\mu}'}\delta(x-x') \ .
\end{equation}
Here
\begin{equation}\label{unpert_mode_en}
  k^2_{\bm{\mu}}=\varepsilon_F-\left(\frac{\pi n}{W}\right)^2-\left(\frac{\pi m}{H}\right)^2
\end{equation}
is the unperturbed mode energy,
\begin{equation}\label{V_mu}
  {\mathcal V}_{\bm{\mu}}(x)=V_{\bm{\mu}\bm{\mu}}(x)+
  \frac{H^2}{12l_{B}^4}\left(1-\frac{6}{\pi^2m^2}\right)
\end{equation}
is the diagonal-in-mode-indices matrix element of the total potential which
includes the disorder-induced part, $V(\mathbf{r})$, and all the
magnetic-field-related terms in l.h.s. of Eq.~\eqref{main_eq},
$l_{B}=\sqrt{\Phi_0/2\pi B}$ is the \emph{total} magnetic length. The term
$V_{\bm{\mu}\bm{\mu}}(x)$ in Eq.~\eqref{V_mu} is the diagonal element of the
mode matrix $\|V_{\bm{\mu}\bm{\nu}}\|$, whose elements are evaluated as
\begin{equation}\label{V_mode}
  V_{\bm{\mu}\bm{\nu}}(x)=\int_S\textit{d}\,\mathbf{r}_{\perp}
  \bra{\mathbf{r}_{\perp};\bm{\mu}}V(\mathbf{r})\ket{\mathbf{r}_{\perp};\bm{\nu}}\
  ,
\end{equation}
integration is over cross-section $S$ of the quantum well \eqref{geom}.
Off-diagonal matrix elements $\hat{{\mathcal U}}_{\bm{\mu}\bm{\nu}}(x)$ in
Eq.~\eqref{mode_eqn} likewise include the disorder- and the
magnetic-field-originated potentials,
\begin{equation}\label{U_munu}
  \hat{{\mathcal U}}_{\bm{\mu}\bm{\nu}}(x)=V_{\bm{\mu}\bm{\nu}}(x)+
  2iH\left(\frac{S_{\bm{\mu}\bm{\nu}}^{(x)}}{l_{x}^2W}-
  \frac{S_{\bm{\mu}\bm{\nu}}^{(y)}}{l_{y}^2}\frac{\partial}{\partial x}\right)+
  C_{\bm{\mu}\bm{\nu}}\frac{H^2}{l_{B}^4}\ .
\end{equation}
Here, partial magnetic lengths $l_i$ (with $i=y,z$) are given by
$l_i^2=\Phi_0/2\pi|B_i|$, $S_{\bm{\mu}\bm{\nu}}^{(i)}$ and
$C_{\bm{\mu}\bm{\nu}}$ are the model-specific numerical coefficients, each of
the order unity,
\begin{subequations}\label{SxyC}
\begin{eqnarray}
\label{Sx}
 S_{\bm{\mu}\bm{\nu}}^{(x)} &=& \varsigma_x \frac{4nn_1}{n^2-n_1^2}
  \sin^2\left[\frac{\pi}{2}(n-n_1)\right]
  \big(1-\delta_{mm_1}\big)\frac{8mm_1}{\pi^2(m^2-m_1^2)^2}
  \sin^2\left[\frac{\pi}{2}(m-m_1)\right]\ ,\\
\label{Sy}
 S_{\bm{\mu}\bm{\nu}}^{(y)} &=& \varsigma_y\delta_{nn_1}
  \big(1-\delta_{mm_1}\big)\frac{8mm_1}{\pi^2(m^2-m_1^2)^2}
  \sin^2\left[\frac{\pi}{2}(m-m_1)\right]\ ,\\
\label{Cpar}
 C_{\bm{\mu}\bm{\nu}} &=& \delta_{nn_1}
  \big(1-\delta_{mm_1}\big)\frac{8mm_1}{\pi^2(m^2-m_1^2)^2}
  \cos^2\left[\frac{\pi}{2}(m-m_1)\right]\ .
\end{eqnarray}
\end{subequations}
In Eqs.~\eqref{SxyC}, $\varsigma_i=B_i/|B_i|=\sgn B_i$ and the notations for
mode indices are employed such that $\bm{\mu}=(n,m)$,
$\bm{\nu}=\nolinebreak(n_1,m_1)$. Note that the specific form of coefficients
\eqref{SxyC} does not significantly affect the algebraic structure of
Eq.~\eqref{mode_eqn} and hence the subsequent transformations. This fact is
the main cause for the final results being uncritically sensitive to the
actual form of the confinement potential.

In Eq.~\eqref{mode_eqn}, matrix elements ${\mathcal V}_{\bm{\mu}}(x)$ and
$\hat{{\mathcal U}}_{\bm{\mu}\bm{\nu}}(x)$ may be regarded as the effective
potentials responsible for intra-mode and inter-mode scattering,
respectively. We thus adopt the approach where interactions of the electron
system with both the disorder potential and the magnetic field are exploited
similarly, i.~e. they are described by the additive static potentials which
are basically different in their correlation properties.

To proceed further with the transverse mode decoupling, we will follow the
procedure outlined in Refs.~\cite{bib:Tar00,bib:Tar03}. It was shown there
that the off-diagonal components of the mode Green matrix are expressed
through corresponding diagonal elements only by means of some linear
operation. Specifically, from Eq.~\eqref{mode_eqn} the operator relation can
be derived,
\begin{equation}\label{GnmGmm}
  \hat{G}_{\bm{\nu}\bm{\mu}}=\bm{P}_{\bm{\nu}}
  (\openone-\hat{\mathsf R})^{-1}\hat{\mathsf R}
  \bm{P}_{\bm{\mu}}\hat{G}_{\bm{\mu}\bm{\mu}}\ ,
\end{equation}
where both $\hat{G}_{\bm{\nu}\bm{\mu}}$ and $\hat{G}_{\bm{\mu}\bm{\mu}}$ are
treated as matrices in extended $x$-coordinate space $\mathbb{X}$. The
operator $\hat{\mathsf R}$ acts in the mixed coordinate-mode space
${\mathsf{\overline M}_{\bm{\mu}}}$ constructed as a direct product of the
space $\mathbb{X}$ and the truncated mode space which incorporates the whole
set of mode indices except the unique index $\bm{\mu}$. In its turn, operator
$\hat{\mathsf R}$ is expressed as a product of operators $\hat{G}^{(V)}$ and
$\hat{\mathcal U}$, i.~e. $\hat{\mathsf R}=\hat{G}^{(V)}\hat{\mathcal U}$,
which are represented in ${\mathsf{\overline M}_{\bm{\mu}}}$ by matrix
elements
\begin{subequations}\label{G(V)U}
\begin{align}\label{GV-matr}
  & \bra{x,\bm{\nu}}\hat{\mathcal G}^{(V)}\ket{x',\bm{\nu}'}=
  G^{(V)}_{\bm{\nu}}(x,x')\delta_{\bm{\nu}\bm{\nu}'}\ ,\\
  \label{U-matr}
  & \bra{x,\bm{\nu}}\hat{\mathcal U}\ket{x',\bm{\nu'}} =
  {\mathcal U}_{\bm{\nu}\bm{\nu'}}(x)\delta(x-x')\ .
\end{align}
\end{subequations}
The function $G^{(V)}_{\bm{\nu}}(x,x')$ in \eqref{GV-matr} is the trial mode
Green function which satisfies the equation resulting from \eqref{mode_eqn}
provided that all inter-mode potentials are put identically equal to zero,
\begin{equation}\label{G_trial}
  \left[\frac{\partial^2}{\partial x^2}+
  k^2_{\bm{\nu}}+i0-{\mathcal V}_{\bm{\nu}}(x)\right]G^{(V)}_{\bm{\nu}}(x,x')=
  \delta(x-x')\ .
\end{equation}
The operator $\bm{P}_{\bm{\nu}}$ in \eqref{GnmGmm} is the projection operator
whose action reduces to assigning the given value $\bm{\nu}$ (or $\bm{\mu}$)
to the nearest mode index of arbitrary operator standing next to it (either
to the left or right), without affecting the product in the $\mathbb{X}$
space. This operator may thus be thought of as contracting the action of
scattering operator $\hat{\mathsf R}$ from three-dimensional space
${\mathsf{\overline M}_{\bm{\mu}}}$ to the unidirectional space $\mathbb{X}$.

Assuming $\bm{\mu}'=\bm{\mu}$ and substituting into Eq.~\eqref{mode_eqn} all
inter-mode propagators in the form \eqref{GnmGmm} we arrive at a closed
equation for the intra-mode (i.~e. mode-diagonal) propagator
$G_{\bm{\mu}\bm{\mu}}(x,x')$, namely
\begin{equation}\label{G_mode-diag}
  \left[\frac{\partial^2}{\partial x^2}+
  k^2_{\bm{\mu}}+i0-
  {\mathcal V}_{\bm{\mu}}(x)-\hat{\mathcal T}_{\bm{\mu}}\right]
  G_{\bm{\mu}\bm{\mu}}(x,x')=\delta(x-x')\ ,
  \qquad\text{for}\ \ \forall\bm{\mu}\ .
\end{equation}
Here,
\begin{equation}\label{T-oper}
  \hat{\mathcal T}_{\bm{\mu}}=\bm{P}_{\bm{\mu}}\hat{\mathcal U}
  (\openone-\hat{\mathsf R})^{-1}\hat{\mathsf R}\bm{P}_{\bm{\mu}}\ ,
\end{equation}
is the operator (in $\mathbb{X}$ space) potential, which in general case is
highly non-local even though the magnetic-field-induced potentials may be
completely non-existent. This potential bears a strong resemblance to the
$T$-matrix which is well-known in quantum theory of scattering
\cite{bib:Newton68,bib:Taylor72}. Normally, this matrix is quite singular in
the multi-channel case. However, in our approach, as was proven in
\cite{bib:Tar00}, the $T$-operator is properly regularized owing to
separation of the intra- and inter-mode potentials in Eq.~\eqref{mode_eqn}.
We will refer below to this operator as the \emph{effective} inter-mode
potential.

Bearing in mind the explicit form \eqref{V_mu} of the potential ${\mathcal
V}_{\bm{\mu}}(x)$, in particular, the non-random nature of its ``magnetic''
part, it is worthwhile to renormalize unperturbed mode energy
\eqref{unpert_mode_en} by introducing, in place of $k^2_{\bm{\mu}}$, the
magnetic-field-modified longitudinal energy
\begin{equation}\label{moden-renorm}
  \varkappa_{\bm{\mu}}^2=k^2_{\bm{\mu}}-
  \frac{H^2}{12l_{B}^4}\left(1-\frac{6}{\pi^2m^2}\right)\ .
\end{equation}
As a result, in place of equations \eqref{G_trial} and \eqref{G_mode-diag} we
are led to analyze a couple of different, though equivalent, equations,
\begin{subequations}\label{Eqs-renorm}
\begin{align}\label{Eqs-renorm-trial}
  & \left[\frac{\partial^2}{\partial x^2}+
  \varkappa^2_{\bm{\nu}}+i0-V_{\bm{\nu}\bm{\nu}}(x)\right]G^{(V)}_{\bm{\nu}}(x,x')=
  \delta(x-x')\ , \\
  \label{Eqs-renorm-main}
  & \left[\frac{\partial^2}{\partial x^2}+
  \varkappa^2_{\bm{\mu}}+i0-
  V_{\bm{\mu}\bm{\mu}}(x)-\hat{\mathcal T}_{\bm{\mu}}\right]
  G_{\bm{\mu}\bm{\mu}}(x,x')=\delta(x-x')\ ,
\end{align}
\end{subequations}
where solely the disorder potential $V_{\bm{\mu}\bm{\mu}}(x)$ appears instead
of the total potential \eqref{V_mu} and the replacement has been performed
$k^2_{\bm{\mu}}\to\varkappa^2_{\bm{\mu}}$.

\section{Spectral peculiarities of renormalized mode states}

Before we proceed to the conductance calculation, let us examine fist
spectral properties of the open quantum system in question, see
Ref.~\cite{bib:Tar04prep} for more details. Although the true eigenstates are
unknown in the presence of a random potential, equation
\eqref{Eqs-renorm-main} permits of analyzing the particular mode state
$\bm{\mu}$ comprehensively. This can be done due to one-dimensional character
of the equation of motion which can be regarded as completely decoupled from
the remaining bath of mode states, whatever strength of the entangling
potentials in Eq.~\eqref{mode_eqn}.

Consider the operator in square brackets of Eq.~\eqref{Eqs-renorm-main},
assuming equation \eqref{Eqs-renorm-trial} to be \emph{ad interim} solved
(see Appendix) and the potential $\hat{\mathcal T}_{\bm{\mu}}$ thus
completely determined. Our aim is to examine the average value of this
operator as it specifies the energy of the particular mode in the mean-field
approximation. This approximation makes practical sense on condition that
scattering produced by fluctuating parts of the potentials may be regarded as
weak. As far as the disorder potential $V_{\bm{\mu}\bm{\mu}}(x)$ is
concerned, the weak scattering (WS) condition is normally cast to
inequalities
\begin{equation}\label{weak_imp}
  k^{-1}_F,\,r_c\ll\ell\ ,
\end{equation}
where $\ell$ stands for the electron mean free path at a zero magnetic field.
For the reference purpose, this path calculated from the model of the
white-noise Gaussian-distributed potential, whose binary correlation function
is $\big<V(\mathbf{r})V(\mathbf{r}')\big>=\mathcal{Q}\delta(\mathbf{r}-
\nolinebreak\mathbf{r}')$, equals $4\pi/\mathcal{Q}$.

Given the external magnetic field, there appears an extra length parameter in
the problem in question, specifically, the magnetic length, in terms of which
it would be advisable to specify the condition where the
magnetic-field-induced scattering could be regarded as weak. In our approach,
this type of scattering is taken into consideration in two fundamentally
different places. The first one is the ``magnetic'' part of the potential
\eqref{V_mu}, which is absorbed in the mode energy renormalization
\eqref{moden-renorm}. The other place is the mode-mixing potential
\eqref{T-oper}. In this study, we will focus on the limiting case of
relatively weak magnetic fields, where maximal cyclotron radius of classical
electron orbit, $R_c~=~k_Fl_B^2$, is large as compared to the quantum well
thickness $H$. It was shown in \cite{bib:Tar04prep} that under condition
\eqref{weak_imp} of weak disorder-related scattering (WDS) in conjunction
with the constraint
\begin{equation}\label{weak_magn}
  \left(\frac{H}{R_c}\right)^2\ll 1\ ,
\end{equation}
which will be hereinafter referred to as the condition for weak magnetic
scattering (WMS ), the norm of the inter-mode scattering operator
$\hat{\mathsf R}$ in \eqref{T-oper} is small compared to unity,
\begin{equation}\label{Rnorm_est}
  \|\hat{\mathsf R}\|^2\sim\frac{1}{k_F\ell}+
  \left(\frac{H}{R_c}\right)^2\ll 1\ .
\end{equation}
This limitation allows the inverse operator in \eqref{T-oper} to be expanded
in series and the operator potential $\hat{\mathcal T}_{\bm{\mu}}$ to be
simplified to the form
\begin{equation}\label{T-approx}
  \hat{\mathcal T}_{\bm{\mu}}\approx
  \bm{P}_{\bm{\mu}}\hat{\mathcal U}
  \hat{\mathcal G}^{(V)}\hat{\mathcal U}\bm{P}_{\bm{\mu}} \ .
\end{equation}

Unlike quasi-local intra-mode potential $V_{\bm{\mu}\bm{\mu}}(x)$, the
potential \eqref{T-approx} possesses a nonzero mean value, even with no
magnetic field. Bearing in mind the further applications of some perturbation
theory, it is advisable to divide this operator into the mean and the
fluctuating parts, $\hat{\mathcal T}_{\bm{\mu}}=\big<\hat{\mathcal
T}_{\bm{\mu}}\big>+\nolinebreak\Delta\hat{\mathcal T}_{\bm{\mu}}$. In view of
definition \eqref{U_munu}, the average operator $\big<\hat{\mathcal
T}_{\bm{\mu}}\big>$ can be split, though quite conventionally, into the sum
of quasi-local ``disorder'' and essntially non-local ``magnetic'' parts, viz.
$\big<\hat{\mathcal T}_{\bm{\mu}}\big>=\big<\hat{\mathcal
T}^{(\mathcal{Q})}_{\bm{\mu}}\big>+ \big<\hat{\mathcal
T}^{(B)}_{\bm{\mu}}\big>$. The action of the operators $\big<\hat{\mathcal
T}^{(\mathcal{Q})}_{\bm{\mu}}\big>$ and $\big<\hat{\mathcal
T}^{(B)}_{\bm{\mu}}\big>$ on the Green function $G_{\bm{\mu}\bm{\mu}}(x,x')$
is specified by equalities
\begin{subequations}\label{T-action}
\begin{align}\label{T-action-Q}
  \Big[\big<\hat{\mathcal
  T}^{(\mathcal{Q})}_{\bm{\mu}}\big>G_{\bm{\mu}\bm{\mu}}\Big](x,x') =&
  \sum_{\bm{\nu}\neq\bm{\mu}}\int_L\textit{d}x_1
  \Big<V_{\bm{\mu}\bm{\nu}}(x)G^{(V)}_{\bm{\nu}}(x,x_1)V_{\bm{\nu}\bm{\mu}}(x_1)\Big>
  G_{\bm{\mu}\bm{\mu}}(x_1,x') \\
\intertext{and}
  \big[\big<\hat{\mathcal
  T}_{\bm{\mu}}^{(B)}\big>G_{\bm{\mu}\bm{\mu}}\big](x,x')=&
  \sum_{\bm{\nu}\neq\bm{\mu}}
  \left[2iH\left(\frac{S_{\bm{\mu}\bm{\nu}}^{(x)}}{Wl_{x}^2}-
  \frac{S_{\bm{\mu}\bm{\nu}}^{(y)}}{l_{y}^2}\frac{\partial}{\partial
  x}\right)+ C_{\bm{\mu}\bm{\nu}}\frac{H^2}{l_{B}^4}\right]
  \notag\\
  \label{T-action-B}
  & \times\int_L\textit{d}x_1\big<G^{(V)}_{\bm{\nu}}(x,x_1)\big>
  \left[2iH\left(\frac{S_{\bm{\bm{\nu}\mu}}^{(x)}}{Wl_{x}^2}-
  \frac{S_{\bm{\nu}\bm{\mu}}^{(y)}}{l_{y}^2}\frac{\partial}{\partial
  x_1}\right)+ C_{\bm{\nu}\bm{\mu}}\frac{H^2}{l_{B}^4}\right]
  G_{\bm{\mu}\bm{\mu}}(x_1,x')\ .
\end{align}
\end{subequations}

To proceed further with expressions \eqref{T-action}, it is necessary to
specify the trial Green function $G^{(V)}_{\bm{\nu}}(x,x')$ or, more
precisely, its averaged value $\big<G^{(V)}_{\bm{\nu}}(x,x')\big>$. In doing
so, we will use a somewhat simplified version of correlation relations for
the disorder potential, namely
\begin{subequations}\label{imp_corr}
\begin{align}
 \label{corr1}
  \big<V(\mathbf{r})\big>&=0\ ,\\
 \label{corr3d_simpl}
  \big<V(\mathbf{r})V(\mathbf{r}')\big>&=\mathcal{QW}(x-x')
  \delta(\mathbf{r}_{\perp}-\mathbf{r}'_{\perp}) \ .
\end{align}
\end{subequations}
As shown in the Appendix, under WS conditions which imply that WDS
inequalities \eqref{weak_imp} and WMS inequality \eqref{weak_magn} are
satisfied simultaneously, the averaging of the trial Green function results
in the asymptotic expression
\begin{equation}\label{G^V_nu}
  \big<G^{(V)}_{\bm{\nu}}(x,x')\big>\approx\frac{-i}{2\varkappa_{\bm{\nu}}}
  \exp\left\{\left[i\varkappa_{\bm{\nu}}-\frac{1}{2}\left(\frac{1}{L_f^{(V)}(\bm{\nu})}+
  \frac{1}{L_b^{(V)}(\bm{\nu})}\right)\right]|x-x'|\right\}\ .
\end{equation}
Here, $L_{f,b}^{(V)}(\bm{\nu})$ are the forward ($f$) and the backward ($b$)
scattering lengths of the trial mode $\bm{\nu}$, which are determined from
model \eqref{imp_corr} as \cite{bib:Tar03}
\begin{equation}\label{ext_length}
  L_f^{(V)}(\bm{\nu})=\frac{4S}{9\mathcal{Q}}
  \left(2\varkappa_{\bm{\nu}}\right)^2 \ ,\hspace{2cm}
  L_b^{(V)}(\bm{\nu})=\frac{4S}{9\mathcal{Q}}
  \frac{\left(2\varkappa_{\bm{\nu}}\right)^2}%
  {\widetilde{\mathcal{W}}(2\varkappa_{\bm{\nu}})} \ ,
\end{equation}
$\widetilde{\mathcal{W}}(q)$ is the Fourier transform of the function
$\mathcal{W}(x)$ from \eqref{corr3d_simpl}. Expression \eqref{G^V_nu} is
literally applicable to extended modes, i.~e. to modes with
$\varkappa_{\bm{\nu}}^2>0$. with $\varkappa_{\bm{\nu}}^2<0$ (evanescent
modes) one should assume $\varkappa_{\bm{\nu}}=i|\varkappa_{\bm{\nu}}|$ in
Eq.~\eqref{G^V_nu} and both of the extinction lengths \eqref{ext_length} to
turn to infinity.

With function \eqref{G^V_nu}, the action of the operator $\big<\hat{\mathcal
T}^{(\mathcal{Q})}_{\bm{\mu}}\big>$ reduces to multiplying the operand in
Eq.~\eqref{T-action-Q} by the complex-valued disorder-related self-energy
factor \cite{bib:Tar00,bib:Tar03}, specifically, $\Big[\big<\hat{\mathcal
T}^{(\mathcal{Q})}_{\bm{\mu}}\big>G_{\bm{\mu}\bm{\mu}}\Big](x,x')
=-\varSigma^{(\mathcal{Q})}_{\bm{\mu}}G_{\bm{\mu}\bm{\mu}}(x,x')$, where
$\varSigma^{(\mathcal{Q})}_{\bm{\mu}}=\Delta\varkappa_{\bm{\mu}}^2+
i/\tau_{\bm{\mu}}^{(\varphi)}$, with
\begin{subequations}\label{renorm_imp_spec}
\begin{align}
\label{ren_en}
  \Delta \varkappa_{\bm{\mu}}^2 &=
  \frac{\mathcal{Q}}{S}\sum_{\bm{\nu}\neq\bm{\mu}}\mathcal{P}
  \int_{-\infty}^{\infty}\frac{\mathrm{d}q}{2\pi} \;
  \frac{\widetilde{\mathcal{W}}(q+\varkappa_{\bm{\mu}})}{q^2-\varkappa_{\bm{\nu}}^2}
  \ , \\
\label{dephase}
  \frac{1}{\tau_{\bm{\mu}}^{(\varphi)}} &=
  \frac{\mathcal{Q}}{4S}
  \overline{\sum_{\bm{\nu}\neq\bm{\mu}}} \frac{1}{\varkappa_{\bm{\nu}}}
  \left[\widetilde{\mathcal{W}}(\varkappa_{\bm{\mu}}-\varkappa_{\bm{\nu}})+
  \widetilde{\mathcal{W}}(\varkappa_{\bm{\mu}}+\varkappa_{\bm{\nu}})\right]
  \ .
\end{align}
\end{subequations}
Symbol $\mathcal{P}$ in Eq.~\eqref{ren_en} stands for the integral principal
value, the bar over the summation index in \eqref{dephase} signifies the
summation over extended modes only, if any.

The conditional character of the term ``disorder self-energy'' with reference
to $\varSigma^{(\mathcal{Q})}_{\bm{\mu}}$ suggests that this factor is
actually determined not solely in terms of the disorder potential, which
results in the pre-factor of $\mathcal{Q}$, but also is influenced by the
magnetic field. The latter field renormalizes wavenumbers
$\varkappa_{\bm{\mu},\bm{\nu}}$ and adjusts the number of extended modes, see
below. Here it should be noted that the dephasing factor \eqref{dephase} can
only be nonzero if quench-disorder-induced scattering is allowed between
different extended modes. Otherwise the coherence of the trial mode states is
unaffected by static disorder.

Apart from the disorder-induced fraction of the mode self-energy, the action
of essentially non-local ``magnetic'' part of the operator
$\big<\hat{\mathcal T}_{\bm{\mu}}\big>$ results in additional self-energy
term which will later on be referred to as magnetic self-energy,
\begin{align}
  \varSigma^{(B)}_{\bm{\mu}}=
  \sum_{\bm{\nu}\neq\bm{\mu}}
  &\left[2iH\left(\frac{S_{\bm{\mu}\bm{\nu}}^{(x)}}{Wl_{x}^2}-
  i\varkappa_{\bm{\mu}}\frac{S_{\bm{\mu}\bm{\nu}}^{(y)}}{l_{y}^2}\right)+
  C_{\bm{\mu}\bm{\nu}}\frac{H^2}{l_{B}^4}\right]\notag\\
\label{B_selfenergy}
  & \times\frac{1}{\varkappa_{\bm{\mu}}^2-\varkappa_{\bm{\nu}}^2-
  i\varkappa_{\bm{\nu}}\left(\frac{1}{L_f^{(V)}(\bm{\nu})}+
  \frac{1}{L_b^{(V)}(\bm{\nu})}\right)}
  \left[2iH\left(\frac{S_{\bm{\nu}\bm{\mu}}^{(x)}}{Wl_{x}^2}-
  i\varkappa_{\bm{\mu}}\frac{S_{\bm{\nu}\bm{\mu}}^{(y)}}{l_{y}^2}\right)+
  C_{\bm{\nu}\bm{\mu}}\frac{H^2}{l_{B}^4}\right]\ .
\end{align}
The principal difference between the disorder and the magnetic self-energies
is that the former takes a non-zero value provided that the disorder strength
(which is specified by the parameter $\mathcal{Q}$) be finite. On the
contrary, the magnetic-field-induced term $\varSigma^{(B)}_{\bm{\mu}}$ only
vanishes in the limit of the magnetic field equal to zero, remaining
otherwise finite at any disorder strength. By comparing imaginary parts of
the terms $\varSigma^{(\mathcal{Q})}_{\bm{\mu}}$ and
$\varSigma^{(B)}_{\bm{\mu}}$ one can determine that the ratio of ``magnetic''
and ``disorder'' dephasing rates is evaluated as
\begin{equation}\label{B/imp}
  \frac{\Im\varSigma^{(B)}_{\bm{\mu}}}{\Im\varSigma^{(\mathcal{Q})}_{\bm{\mu}}}\sim
  \left[\frac{H}{R_ck_FW}\left(\frac{B_x}{B}\right)^2+
  \frac{H}{R_c}\left(\frac{B_y}{B}\right)^2+\frac{H^2}{R_c^2}\right]^2\ll 1\ .
\end{equation}
It is clear that under WMS condition \eqref{weak_magn} the
magnetic-field-originated dephasing is negligible, whatever level of the
disorder. One is thus led to conclude that strong inter-mode mixing resulting
from the magnetic field only does not cause the mode energy levels to be
considerably widened unless there exists a \emph{random} potential due to
some kind of disorder serving as a mediator for the magnetic-field-related
dephasing effect. As far as mode entanglement is concerned, it is evident
from \eqref{renorm_imp_spec} that the specific role of the magnetic field is
to change the collective parameters of the electron spectrum, such as the
mode content of the confined system of carriers and the mode density of
states (MDOS). It is only in this indirect manner that the mode level width
becomes dependent on the magnetic field.

The number of extended modes in the quantum waveguide, which substantially
governs dephasing rate \eqref{dephase}, is specified by the requirement for
mode energies to be positive-valued. Prior to analyzing in WS approximation
the role of fluctuating potentials $V_{\bm{\mu}\bm{\mu}}(x)$ and
$\Delta\hat{\mathcal T}_{\bm{\mu}}$ one can notice that the energy of a
specific mode depends not only on the confining potential but also on the
magnetic field that likewise produces the confinement effect. As seen from
\eqref{moden-renorm}, the new ``unperturbed'' mode energy
$\varkappa_{\bm{\mu}}^2$ decreases steadily with the magnetic field growth,
whatever the orientation within the plane of a~2D system. A decrease in that
energy should result in truncation of the number of extended modes
responsible for current transport and, therefore, in the conductance
abruptly falling off. The latter effect is quite often interpreted in terms
of electron effective mass enhancement \cite{bib:ADPGTV04}.

Although magnetic renormalization \eqref{moden-renorm} is of purely
intra-mode nature, corrections of the same sign to the mode energy are
produced by the inter-mode magnetic scattering as well, which is taken into
account by self-energy \eqref{B_selfenergy}. This is indeed the case as long
as WMS condition holds true. Moreover, inter-mode magnetic scattering can
lead to some anisotropy of the electron effective mass with respect to the
magnetic field in-plane orientation, depending on symmetry properties of
numerical coefficients in Eq.~\eqref{B_selfenergy}, i.~e. on geometrical
symmetry of the quantum well. Note that the intra-mode magnetic term in
\eqref{moden-renorm} is entirely isotropic.

As the magnetic field increases, the terms in square brackets of
\eqref{B_selfenergy}, which are proportional to $C_{\bm{\mu}\bm{\nu}}$, will
exceed the rest of the terms, thus resulting in the subsequent increase of
the mode energy. This effect, however, corresponds to the range of relatively
strong magnetic fields, where WMS inequality \eqref{weak_magn} is violated.
In this paper, this case will not be dealt with.

\begin{figure}[h]
\centering
\setcaptionmargin{1in}%
{\includegraphics[width=10cm]{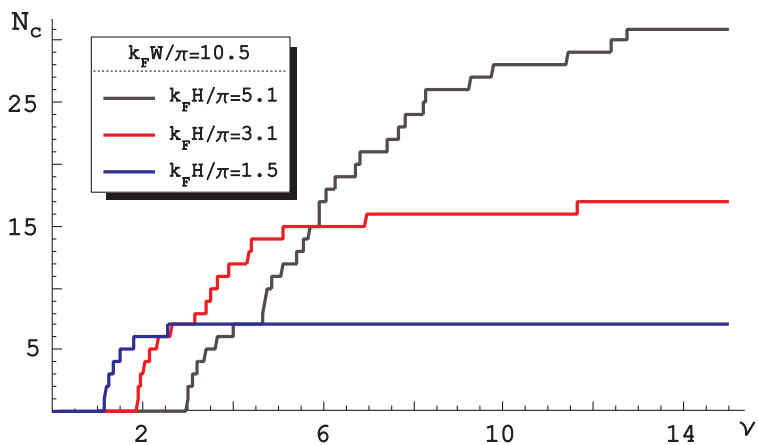}}%
\captionstyle{hang}%
\caption{The dependence of the number of extended modes in the quantum
waveguide on the inverse magnetic field ($\nu$ is the Landau filling
factor).\hfill\label{fig2}}
\end{figure}

In Figure~\ref{fig2}, the numerical dependence of the extended mode number
$N_c$, commonly referred to as the number of effective conducting channels,
is reproduced from Ref.~\cite{bib:Tar04prep}. The magnetic field assumed to
be codirectional with the axis of current flow is scaled as the Landau-level
filling factor $\nu=\nolinebreak (k_Fl_B)^2=\nolinebreak k_FR_c$. The
collapse of the number of current-carrying modes with a growing magnetic
field is apparent, regardless of the quantum well thickness $H$, the width
$W$ being held constant. Numerical considerations reveal that in-plane
rotation of the magnetic field slightly changes the picture presented. This
is consistent with the fact that the real part of self-energy
\eqref{B_selfenergy} can reach, at most, the same value (on the order of
magnitude) as the magnetic term in \eqref{moden-renorm}, and is in agreement
with experiments~\cite{bib:PBPB02}.

\begin{figure}[h]
\centering
\setcaptionmargin{1in}%
{\includegraphics[width=10cm]{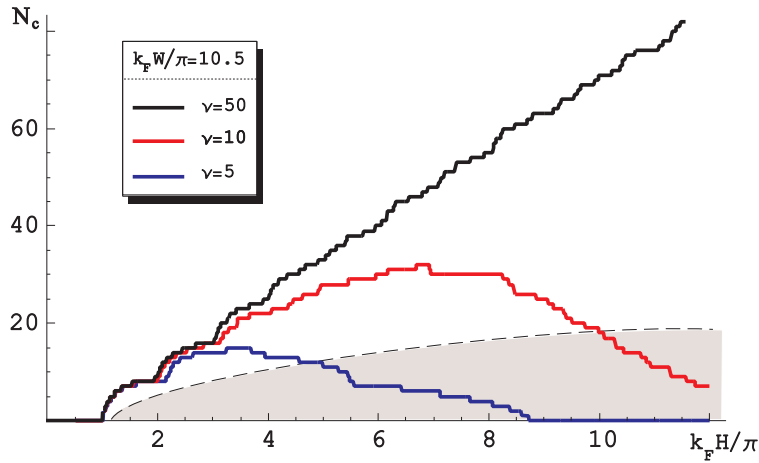}}%
\captionstyle{hang}%
\caption{The number of conducting channels vs the width of the near-surface
potential well at different values of the Landau filling factor. The shaded
area below the dashed curve covers the parameter region where the WMS
criterion is violated.\hfill\label{fig3}}
\end{figure}

We also reproduce from \cite{bib:Tar04prep} the $N_c$ dependence on the
effective thickness of the quantum waveguide, the latter adjusted in practice
by depletion voltage. In the lowest magnetic fields (upper curve in
Fig.~\ref{fig3}) the number of channels shows a near-linear increase with
growing $H$. This corresponds to standard geometrical considerations
applicable to systems of waveguide configuration. As the magnetic field is
getting higher,
the conventional geometric increase in the number of channels slows down,
gradually changing to a decrease in the conducting mode number. This unusual
non-monotonicity of the mode content of the quantum system is due to
essentially non-monotonic dependence on $H$ of the
magnetic-field-renormalized mode energy \eqref{moden-renorm}.

The parallel magnetic field effect on the number of subbands in a quantum
well was previously noticed by some authors \cite{bib:SdS89,bib:SdS93} with
reference to semiconductor heterostructures. The reduction in the number of
conducting channels with a growth in the magnetic field is certainly
a~quantum-mechanical effect pertinent to multi-electron systems. As a matter
of fact, its nature is closely related to the Aharonov-Bohm (AB) phase
incursion, which is known to effectively change carriers' energy
\cite{bib:Olariu85,bib:Wash91} and has an impact on the interference patterns
observed in experiments with AB rings. In the problem under study, unlike
quasi-1D metal rings subjected to magnetic field, the AB phase is essentially
different for electrons following different orbits. Therefore, it may seem to
be difficult to take the magnetic field into account by conventional wave
function phase renormalization. Yet this phase effect appears in the
\emph{collective} mode energies, affecting crucially the number of
current-carrying modes in a~laterally confined electron system.

\begin{figure}[h]
\centering
\setcaptionmargin{1in}
{\includegraphics[width=10cm]{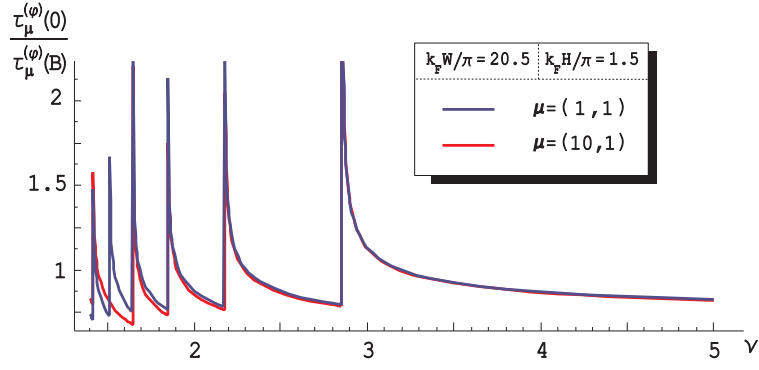}} \captionstyle{hang}\caption{The
dephasing rate \eqref{dephase} for two particular modes vs inverse magnetic
field. \hfill\label{fig4}}
\end{figure}

The apparent impact of the in-plane magnetic field on the number of extended
modes inevitably occurs in the magnetic-field dependence of the mode
dephasing rate. While in the limit of a large number of conducting channels
the replacement of the sum in Eq.~\eqref{dephase} with the integral leads to
the familiar quasi-classical formula,
\begin{equation}\label{deph(B)}
  \frac{1}{\tau_{\bm{\mu}}^{(\varphi)}(B)}\approx
  \frac{1}{\tau_{\bm{\mu}}^{(\varphi)}(0)}\sqrt{1-
  \frac{H^2}{12R_c^2}}\ ,
\end{equation}
where $1/\tau_{\bm{\mu}}^{(\varphi)}(0)=k_F\mathcal{Q}/4\pi$ stands for the
dephasing rate due to disorder scattering in the absence of magnetic
field~\cite{bib:Tar03}, in Fig.~\ref{fig4} we give the numerically obtained
dependence on the magnetic field of two particular modes of the quantum
waveguide under study. Well-known van~Hove singularities of MDOS, which are
directly related to step-wise change in the number of extended modes, are
well pronounced on both of the curves.

\begin{figure}[h]
\centering
\setcaptionmargin{1in}
{\includegraphics[width=10cm]{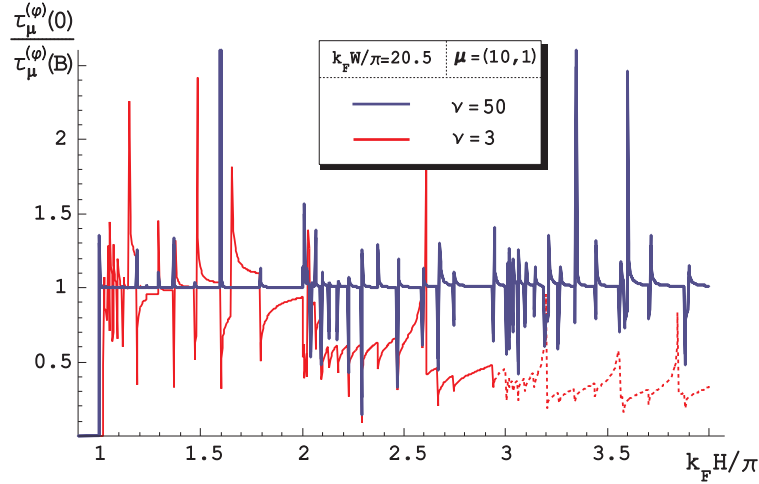}} \captionstyle{hang}\caption{The
dephasing rate vs the quantum waveguide thickness at different strengths of
in-plane magnetic field. The broken fraction of the lower curve falls into
the range of parameters where WMS condition \eqref{weak_magn} is violated.
\hfill\label{fig5}}
\end{figure}

Besides, Fig.~\ref{fig5} illustrates the dephasing rate of the particular
mode vs variable effective thickness of Q2D system of carriers. Availability
of van~Hove singularities similar to those depicted in Fig.~\ref{fig4}
compels one to conclude that by means of the orbital coupling to Q2D
electrons the in-plane magnetic field has the effect which is, in a way,
similar to that of the electrostatic confinement potential. At the same time,
in contrast to the magnetic-field-controlled singularities shown in
Fig.~\ref{fig4}, the oscillations of truly geometrical origin are noticeably
more complicated. The distinction is attributed to a substantially different
response of the effective mode energy \eqref{moden-renorm} to the magnetic
field, on the one hand, and to size parameters of the confined electron
system, on the other. However, it should be emphasized that in both of the
graphs, \ref{fig4}~and~\ref{fig5}, the reduction of the dephasing by the
quenched disorder is clearly visible as the magnetic field grows. This fact
is indicative of an increase in coherence of the electron transport in
quench-disordered Q2D systems subjected to an external magnetic field.

\section{Magnetoconductance of a Q2D system}

By substituting the Green functions into Eq.~\eqref{Kubo_cond} in the form of
expansion in series over transverse Hamiltonian eigenfunctions we obtain the
following mode representation of the conductance,
\begin{align}
  g_{xx}(L,\mathbf{B})=\frac{2}{L^2}\iint_L\textit{d}x\textit{d}x'
  \sum_{\ \genfrac{}{}{0pt}{}{\bm{\nu}_1,\bm{\nu}_2}{\bm{\nu}_3,\bm{\nu}_4}}
  &\left(\frac{\partial}{\partial x}\delta_{\bm{\nu}_1\bm{\nu}_2}
  -i\varsigma_y\frac{\hat{z}_{\bm{\nu}_1\bm{\nu}_2}}{l_y^2}\right)
  \Big[G^{a}_{\bm{\nu}_2\bm{\nu}_3}(x,x')-G^{r}_{\bm{\nu}_2\bm{\nu}_3}(x,x')\Big]\notag\\
  \label{cond_xx-mode}
  \times &\left(\frac{\partial}{\partial x'}\delta_{\bm{\nu}_3\bm{\nu}_4}
  -i\varsigma_y\frac{\hat{z}_{\bm{\nu}_3\bm{\nu}_4}}{l_y^2}\right)
  \Big[G^{a}_{\bm{\nu}_4\bm{\nu}_1}(x',x)-G^{r}_{\bm{\nu}_4\bm{\nu}_1}(x',x)\Big]\ .
\end{align}
Here, $\delta_{\bm{\nu}_i\bm{\nu}_j}$ is the vector-argument Kronecker delta,
$\hat{z}_{\bm{\nu}_i\bm{\nu}_j}$ is the inter-mode matrix element of the
$z$-coordinate operator which, given the model of the confining potential,
assumes the value
\begin{align}\label{z_nunu1}
  \hat{z}_{\bm{\nu}\bm{\nu}'}=
  -H\delta_{nn'}(1-\delta_{mm'})\frac{8mm'}{\pi^2\left(m^2-m'^2\right)^2}
  \sin^2\left[\frac{\pi}{2}(m-m')\right]\ ,
  \qquad\qquad\bm{\nu}=(n,m)\ .
\end{align}
Using interrelation \eqref{GnmGmm} between off-diagonal and diagonal mode
propagators, and estimate \eqref{Rnorm_est} for the inter-mode scattering
operator it is easy to make sure that the principal contribution to
\eqref{cond_xx-mode} arises from the terms whose mode indices are coincident
with one another. As a result, the conductance is expressed, asymptotically
in WS limit, as
\begin{equation}\label{g_xx-main}
  g_{xx}(L,\mathbf{B})\approx\frac{2}{L^2}
  \sum_{\bm{\nu}}\iint_L\textit{d}x\textit{d}x'
  \frac{\partial}{\partial x}
  \Big[G^{a}_{\bm{\nu}\bm{\nu}}(x,x')-G^{r}_{\bm{\nu}\bm{\nu}}(x,x')\Big]
  \frac{\partial}{\partial x'}
  \Big[G^{a}_{\bm{\nu}\bm{\nu}}(x',x)-G^{r}_{\bm{\nu}\bm{\nu}}(x',x)\Big]\ .
\end{equation}
This formula, in conjunction with \eqref{Eqs-renorm}, ultimately reduces the
initially stated three-dimensional dynamic problem to a set of decoupled,
strictly 1D problems which in the general case are non-Hermitian.

Under WS conditions, equation \eqref{Eqs-renorm-main} can be solved
analytically with an arbitrary accuracy in fluctuating potentials. However,
there is no need to overestimate the expected result. In
Refs.~\cite{bib:Tar00,bib:Tar03}, it was proven that for inactive media with
channel number $N_c>1$ the presence of dephasing term \eqref{dephase} in the
operator potential \eqref{T-approx} enables one to neglect fluctuating
potentials in Eq.~\eqref{Eqs-renorm-main}, thereby seeking intra-mode
propagators from the deterministic equation
\begin{equation}\label{Gmm(0)-eq}
  \left(\frac{\partial^2}{\partial x^2}+
  \varkappa^2_{\bm{\mu}}+
  i/\tau_{\bm{\mu}}^{(\varphi)}\right)
  G_{\bm{\mu}\bm{\mu}}(x,x')=\delta(x-x')\ .
\end{equation}
The solution to this equation, which obeys the openness conditions at the
ends of the interval $\mathcal{L}=(-L/2,L/2)$, under WS conditions is written
as
\begin{equation}\label{Gmm(0)-sol}
  G_{\bm{\mu}\bm{\mu}}(x,x')\approx
  \frac{1}{2i\varkappa_{\bm{\mu}}}\exp\left[\left(i\varkappa_{\bm{\mu}}-
  1/l_{\bm{\mu}}^{(\varphi)}\right)|x-x'|\right]\ ,
\end{equation}
where $l_{\bm{\mu}}^{(\varphi)}=
2\varkappa_{\bm{\mu}}\tau_{\bm{\mu}}^{(\varphi)}$ is the length of the mode
$\bm{\mu}$ phase coherence. With function \eqref{Gmm(0)-sol} substituted into
Eq.~\eqref{g_xx-main}, the~average magnetoconductance is given as
\begin{equation}\label{gxx-res}
  \big<g_{xx}(L,\mathbf{B})\big>=
  \overline{\sum_{\bm{\nu}\neq\bm{\mu}}}
  \frac{l_{\bm{\mu}}^{(\varphi)}}{L}\left[1-\frac{l_{\bm{\mu}}^{(\varphi)}}{L}
  \exp\left(-\frac{L}{l_{\bm{\mu}}^{(\varphi)}}\right)
  \sinh\frac{L}{l_{\bm{\mu}}^{(\varphi)}}\right]\ .
\end{equation}
Here, the terms corresponding to extended modes only are kept since the
evanescent-mode Green functions, which are strongly localized in
$x$-direction, are real-valued and cancel each other in
Eq.~\eqref{g_xx-main}. Note that formula~\eqref{gxx-res} can be viewed as
describing the system conductance in the presence of both the random
scatterers and the magnetic field, which enters implicitly through mode
coherence lengths.

From general result \eqref{gxx-res}, conventional limiting formulae for the
conductance can readily be obtained. In particular, in ballistic limit
$L\ll\ell$ the dimensionless conductance becomes nearly equal to the number
of open conducting channels,
\begin{equation}\label{g_xx-ball}
  \big<g_{xx}(L,\mathbf{B})\big>\approx N_c(\mathbf{B})\ .
\end{equation}
This number, as it follows from the above analysis, is determined by both the
geometric confinement of the electron system and the magnetic field, see
Figs.~\ref{fig2} and~\ref{fig3}. The perfect system conductance is thus
expected to run in steps as a~function of either depletion voltage or the
value of the in-plane magnetic field.

In diffusion limit $L\gg\ell$, if the potential well in the $z$-direction is
wide enough to contain the large number of quantization levels in this
direction, by replacing the sum in r.h.s. of \eqref{gxx-res} with the
integral we arrive at
\begin{equation}\label{g_xx-diff}
  \big<g_{xx}(L,\mathbf{B})\big>\approx\frac 43\frac{N_c(\mathbf{B})\ell}{L}
  \approx\big<g_{xx}(L,0)\big>\left(1-\frac{H^2}{12R_c^2}\right)\ .
\end{equation}
In the case of the zero magnetic field this result is coincident in form with
classical Drude conductance \cite{bib:Tar03}. If $\mathbf{B}\neq 0$, the
magnetoconductance is negative, being varied smoothly with the magnetic
field, specifically, under the quadratic law. This is because the van~Hove
singularities of MDOS prove to be integrated out in such a~rough calculation.

\begin{figure}[h]
\centering
\setcaptionmargin{1in}%
{\includegraphics[width=10cm]{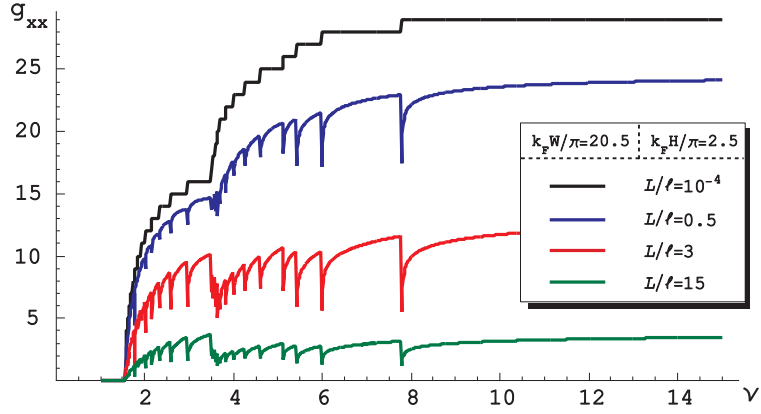}} \captionstyle{hang}\caption{The
conductance vs inverse magnetic field at different disorder level in the
quantum waveguide. \hfill\label{fig6}}
\end{figure}

However, these singularities are actually contained in the mode coherence
lengths, as they are determined using the dephasing rate
formula~\eqref{dephase}. They should appear both in Shubnikov-de~Haas (i.~e.
magnetic-field-driven) oscillations of the conductance and in the conductance
dependence on the quantum well width, which is normally tuned by depletion
voltage. In Fig.~\ref{fig6}, the results numerically obtained from
Eq.~\eqref{gxx-res} at
several values of the diffusion parameter, $L/\ell$, are presented. MDOS
singularities reveal themselves in the form of sharp dips placed close to
those points where the number of conducting modes undergoes stepwise
variations, i.~e. to the thresholds of the transverse subbands, no matter
what the disorder level may be. The disorder seems to manifest itself through
the absolute value of the conductance. Note that in the case of relatively
large disorder (large values of $L/\ell$) the conductance develops
non-monotonically vs the magnetic field, even if MDOS singularities are
smoothed out. When decreasing the mean free path, non-monotonicity becomes so
apparent that it appears to be inadmissible to disregard this effect in the
experimental data. In particular, with regard to this analysis it would be
tempting to revise the observed positive magnetoresistance which is
frequently attributed to spin properties of 2D
systems~\cite{bib:ZMMCG02,bib:MinGerRSGZW03}.

\begin{figure}[h]
\centering
{\includegraphics[width=16cm]{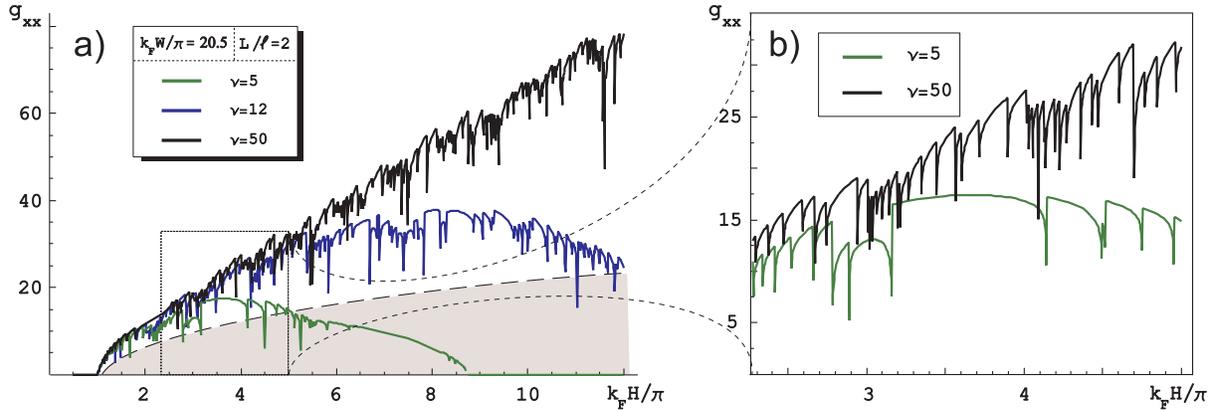}}%
\captionstyle{hang}%
\caption{The conductance vs the width of the near-surface potential well at
different values of the in-plane magnetic field. In panel (b), the region
selected by the rectangle in panel (a), with only two curves left, is zoomed
in to make the dips arisen due to MDOS singularities more clearly visible.
Similarly to Fig.~\ref{fig3}, the shading below the dashed curve signifies
the region where criterion \eqref{weak_magn} is violated. \hfill\label{fig7}}
\end{figure}

To make a comparison with the magnetic-field run, in Fig.~\ref{fig7} the
conductance is shown versus the width of the potential well forming
a~quasi-2D quantum waveguide. Here, MDOS singularities are equally more
pronounced, which demonstrates the change in the number of conducting
channels. Meanwhile, in the latter case these singularities are completely
anticipated from the very outset, since the number of channels is normally
associated with size quantization. The peculiar feature should be noted in
Fig.~\ref{fig7} as against the dependence on the magnetic field. The curves
in Fig.~\ref{fig7} tend to become non-monotonic, on average, as the magnetic
field grows. Evidently, this non-monotonicity accounts for the non-monotonic
dependence on $H$ of the mode eigen-energy \eqref{moden-renorm}.

It is instructive to dwell upon the physical nature of dips in
Figs.~\ref{fig6} and~\ref{fig7}. All of them are positioned in the vicinity
of the points corresponding to opening/closing of the conducting channels.
For one thing, as the electron waveguide gets thin or the magnetic field
grows, the closing of the channel must result in a bench-like fall of the
conductance since each of the conducting channels is expected to bring in
exactly one conductance quantum. At the critical point, the edge extended
mode is transformed to the evanescent one, which is localized at a scale of
the mode wavelength and in this way carry no current in an infinitely long
system. Indeed, this picture is demonstrated by the upper, ``ballistic'',
curve in Fig.~\ref{fig6}.

For another, in approaching the transformation point (subband threshold) the
MDOS of the edge mode diverges whereas its mode velocity tends to zero. The
capacious and slow edge mode serves as a destructive sink for dynamic
electrons, leading to a decrease in the conductance. It is clear that the
dips can arise only when we deal with an imperfect system of carriers, where
scattering is allowed from all remaining extended modes to the slow critical
mode. The conductance in the bottom of the dip may thus reach,
hyperbolically, a nearly zero value, which can hardly be grasped in real
experiments because of various unaccounted extra factors.

\section{Discussion and concluding remarks}

We have demonstrated that the appreciable localizing effect produced by
relatively weak in-plane magnetic field on 2D electron and hole systems can
be reasonably interpreted in the context of Fermi liquid theory applied to
spinless electrons which reside in an open near-surface potential well of
finite rather than zero width. For a relatively weak magnetic field, its
coupling to the carrier orbital degree of freedom, even though it is rather
weak from semi-classical point of view, proves to have substantial influence
on the carrier spectrum and, hence, on the conductance.

The conclusion about strong sensitivity of the carrier spectrum to the
in-plane magnetic field is made from the analogy of tightly gated solid-state
systems to classical wave-guiding systems of planar, though
three-dimensional, configuration. The mode content of these well-known
objects is quite sensitive to the anisotropy in their cross-section, that is
to the applied gate voltage as far as electron devices are concerned. A
remarkable feature of the latter type of systems is that in case the magnetic
field is applied to the finite length, these systems can be thought of as
being subjected to both the disorder potential, whose correlation length can
be arbitrary, and the additive deterministic strongly non-local ``magnetic''
potential barrier. Scattering parameters of this barrier are specified by the
magnetic field strength and orientation, on the one hand, and by the length
of magnetically biased section of the quantum waveguide, on the other. The
effect produced by this barrier results from the mismatching of electron
spectra in the inner and outer parts of the quantum well, the inner part
representing the~finite-length electron system under consideration.

If the electron waveguide cross-section is strongly anisotropic, the bulk of
transverse modes in the electron spectrum proves to be efficiently
transformed from extended to evanescent type as the magnetic field grows
slightly. This is because electron scattering from side boundaries of the
confining potential well is, in the strict sense, specular if the boundaries
are considered as completely inhibiting the transverse current flow. In view
of this fact, it seems to be erroneous to assess the magnetic field effect on
the electron system by simply estimating the variation of the electron
trajectory portions between successive collisions with quantum well
boundaries. It can be easily verified from Eqs.\eqref{unpert_mode_en}
and~\eqref{moden-renorm} that the mode truncation effect of the magnetic
field is the more significant the larger is the aspect ratio of the waveguide
cross-section, given the dimension $H$. For $H$ being small enough, such that
only modes with quantization parameter $m=1$ in the corresponding direction
can be regarded as extended, the total number of extended modes in the
quantum waveguide is mainly determined by the larger cross-section dimension,
$W$ in the case under study. Owing to this, even slight alteration of the
magnetic field can transform a considerable number of modes from extended to
evanescent type, thus leading to significant reduction of the conductance,
even though it might have a~large (ballistic) value in the absence of the
magnetic field.

The mode truncation effect of in-plane magnetic field appears to be quite
similar to that of truly geometric confinement of the electron system. The
closing of each of the current-carrying modes, which makes itself evident in
the form of conductance jumps by exactly one conductance quantum in a perfect
system at a zero temperature, should be regarded as a true quantum phase
transition \cite{bib:SGCS97}. This statement is substantiated by the
indisputable fact that there exists a well-defined correlation length in the
vicinity of a~closing point, whose role is played by the wave length of the
marginal extended mode, which tends to diverge as the critical point is
approached. The closing of the last conducting mode by means of the in-plane
magnetic field may thus be regarded as the magnetic-field-driven MIT.

It should be noted that in the proximity to the MIT the conductance is not
quite precisely described by the present theory, since it is hard to satisfy
WMS conditions over the corresponding range of system parameters. At the same
time, closer examination of magnetic-field-originated items in
Eq.~\eqref{Eqs-renorm-main} reveals that the above-described mode truncation
effect and, hence, the very fact of the existence of magnetic-field-driven
MIT, is robust.

In this paper we have not focused our attention upon the conductance
anisotropy with respect to in-plane orientation of the magnetic field. The
corresponding analysis is straightforward. It consists in comparing the real
part of anisotropic self-energy \eqref{B_selfenergy} and the magnetic part of
the trial mode energy \eqref{moden-renorm}. Under WMS condition
\eqref{weak_magn}, both of these energy items are of the same order of
magnitude, save the case of the magnetic field oriented nearly parallel to
the direction of current. We thus expect the conductance anisotropy to be
expressed, at most, moderately. Moreover, in view of numerical coefficients
in Eq.~\eqref{B_selfenergy} being model-specific, the conductance anisotropy
in practice should be governed substantially by the confinement potential
profile. It should be noted that assuming the magnetic field to grow beyond
WMS limit, the isotropic terms proportional to $l_B^{-4}$ can be made
dominating in self-energy \eqref{B_selfenergy}, thereby removing the in-plane
anisotropy of the conductance. This range of the magnetic field strength
needs to be given a special consideration because the ``magnetic
scattering'', which is an important aspect of our approach, cannot be
regarded as weak in this particular case.

Yet another relevant remark should be made concerning the model of the
confinement potential adopted in this study. In some papers where laterally
confined electron systems are dealt with (see, e.~g., Ref.\cite{bib:DSHw00}),
this potential is taken as a~quadratic function. Thus modelled confinement
seems to be beneficial from the technical point of view, as it enables one to
account for the magnetic field non-perturbatively, the corresponding
transverse eigen-states being known as Fock-Darwin levels
\cite{bib:SdS89,bib:SdS93}. It is natural that the precisely zero width of
those levels in the absence of any disorder implies the entire lack of
dephasing due to the magnetic field only. In this case, the
magnetoconductance should not exhibit a dip structure because the latter
results from MDOS singularities arising exclusively in the presence of the
disorder.

In the domain of weak magnetic fields (in a sense of inequality
\eqref{weak_magn}) the quadratic confinement can hardly be substantiated.
Therefore, the issue of the magnetic-field-induced mode entanglement and the
related dephasing of the mode states might seem to be quite topical in this
limiting case. Note, however, that although rectangular confinement possesses
appreciable inter-mode magnetic scattering, it proves not to result in
widening the transverse quantization levels unless some kind of disorder is
also taken into account. The exact form of transverse eigen-functions is of
no fundamental significance for the development of transport theories in mode
representation, but the mere fact of transverse energy quantization does
matter. This prompts us to expect that the magnetic field alone cannot give
rise to noticeable decoherence of electron states for any hard-wall model of
the electron confinement. The specific form of the confinement potential can
only alter the arrangement of transverse quantization levels and,
consequently, influence the coherence properties of the electron system
indirectly. Of primary value is the dephasing resulting from scattering
caused by some kind of \emph{random} (i.~e., in a~sense, uncontrolled)
potential, no matter static or variable it may be.

\begin{acknowledgments}
  This work was partially supported by the Ukrainian Academy of
  sciences, grant No. 12/04--H under the program ``Nanostructure
  systems, nanomaterials and nanotechnologies''.
\end{acknowledgments}

\appendix*

\section{Trial mode Green function in the presence of magnetic field}
\label{trialGF}

\subsection{Reduction of the boundary-value problem to auxiliary
Cauchi problems}

Prior to obtaining the proper boundary conditions for
Eq.~\eqref{Eqs-renorm-trial}, which would conform to the physical definition
of the system openness at the end points $x=\pm L/2$, note that the problem
governed by this equation is absolutely identical to that of
point-source-radiated classical waves in an 1D random medium. Bearing this in
mind, we will extend the driving equation from the finite interval
$\mathcal{L}$ to the whole $x$-axis, writing it down in symbolic form
\begin{equation}\label{model_eq}
  \left[\frac{d^2}{dx^2}+k^2+i0-V(x)\right]G^{(V)}(x,x')=\delta(x-x')\ ,
\end{equation}
where the potential $V(x)$ is assumed to be the finite-support function which
consists of two fundamentally different items. The regular component of this
potential is $V^{(reg)}(x)=\frac{H^2}{12l_{B}^4}
\left(1-\frac{6}{\pi^2m^2}\right)\theta(L/2-|x|)$, with $\theta(x)$ being the
Heaviside unit-step function, whereas the additional random disorder term is
$V^{(dis)}(x)=V_{\bm{\mu}\bm{\mu}}(x)\theta(L/2-|x|)$. For the sake of
clarity we will omit mode index $\bm{\mu}$, implying that both the wave
number $k$ in \eqref{model_eq} and $\varkappa$ below are related exactly to
this mode.

Since there is a need to perform configurational averaging over the potential
$V^{(dis)}(x)$ it is worthwhile to express the solution to
Eq.~\eqref{model_eq} in terms of wave functions of causal type rather than
functions that comply with the initially stated boundary-value (BV) problem.
This is achieved by employing the formula
\begin{equation}\label{Green-Cochi}
  G^{(V)}(x,x')=\frac{1}{\mathcal{W}}
  \big[\psi_+(x)\psi_-(x')\theta(x-x') +
  \psi_+(x')\psi_-(x)\theta(x'-x) \big] \ ,
\end{equation}
where $\psi_{\pm}(x)$ are two different solutions of homogeneous equation
\eqref{model_eq} with the boundary conditions specified for each of them at
only one end of the coordinate axis, viz. $x\to\pm\infty$, depending on the
sign index, $\mathcal{W}$~is the Wronskian of the above solutions. With this
representation, the trial propagator itself satisfies, as it must, the
initial BV problem and the averaging procedure can be accomplished
mathematically correctly, with due account of multiple backscattering.

Inasmuch as the support of the potential in Eq.~\eqref{model_eq} is bounded,
functions $\psi_{\pm}(x)$ may be sought in the form
\begin{subequations}\label{psi_pm}
\begin{align}\label{psi_pm-in}
  \psi_{\pm}(x)&=\pi_{\pm}(x)\mathrm{e}^{i\varkappa (\pm x - L/2)}-
  i\gamma_{\pm}(x)\mathrm{e}^{-i\varkappa (\pm x - L/2)}\hspace{-2cm}
  &\text{at}\qquad |x|\leqslant L/2\ , \\
 \label{psi_pm-out}
  \psi_{\pm}(x)&=c_{\pm}\mathrm{e}^{ik(\pm x-L/2)}\hspace{-2cm}
  &\text{at}\qquad |x|>L/2\ ,
\end{align}
\end{subequations}
where $\varkappa^2=k^2-V^{(reg)}$. Under WDS condition \eqref{weak_imp},
envelope functions $\pi_{\pm}(x)$ and $\gamma_{\pm}(x)$ in \eqref{psi_pm-in}
can be regarded as smooth factors as compared with near-standing fast
exponentials, which leads to the following coupled dynamic equations,
\begin{subequations}\label{dyn-eq_pi-gamma}
\begin{eqnarray}
 \label{dyn-eq_pi-gamma-1}
  && \pm\pi'_{\pm}(x)+
  i\eta(x)\pi_{\pm}(x)+\zeta^*_{\pm}(x)\gamma_{\pm}(x)=0\ ,\\
 \label{dyn-eq_pi-gamma-2}
  && \pm\gamma'_{\pm}(x)-
  i\eta(x)\gamma_{\pm}(x)+\zeta_{\pm}(x)\pi_{\pm}(x)=0\ .
\end{eqnarray}
\end{subequations}
Random functions $\eta(x)$ and $\zeta_{\pm}(x)$ in \eqref{dyn-eq_pi-gamma}
are constructed as narrow packets of spatial harmonics of the disorder
potential,
\begin{subequations}\label{eff_fields}
\begin{align}
 \label{eff_fields-eta}
  & \eta(x)=\frac{1}{2\varkappa}\int_{x-l}^{x+l}\frac{\mathrm{d}t}{2l}
  V^{(dis)}(t)\ ,\\
 \label{eff_fields-zeta}
  & \zeta_{\pm}(x)=\frac{1}{2\varkappa}\int_{x-l}^{x+l}\frac{\mathrm{d}t}{2l}
  V^{(dis)}(t)\exp[2i\varkappa(\pm x-L/2)]\ ,
\end{align}
\end{subequations}
where spatial averaging is over the interval $2l$ of an arbitrary length
intermediate between ``small'' length scales $\varkappa^{-1}$ and $r_c$, on
the one hand, and the large scattering length (to be determined
self-consistently), on the other. These limitations ensure smoothed random
``potentials'' $\eta(x)$ and $\zeta_{\pm}(x)$ to provide the harmonics
$\pm\varkappa$ forward and backward scattering, respectively.

By joining the solutions \eqref{psi_pm-in} and \eqref{psi_pm-out} at the end
points of interval $\mathcal{L}$ we arrive at the necessary BC for envelopes
$\pi_{\pm}(x)$ and $\gamma_{\pm}(x)$, viz.
\begin{subequations}\label{bound_cond}
\begin{align} \label{bc-pi}
  \pi_{\pm}(\pm L/2)= & \mathrm{const}\ ,\\
  \label{bc-gamma}
  \gamma_{\pm}(\pm L/2)= &
  \EuScript{R}^{(B)}\pi_{\pm}(\pm L/2)\ .
\end{align}
\end{subequations}
The quantity
\begin{equation}\label{magn_refl}
  \EuScript{R}^{(B)}=-i\frac{k-\varkappa}{k+\varkappa}\ ,
\end{equation}
as it follows from \eqref{psi_pm-in}, is the amplitude reflection coefficient
from the interface between the magnetically biased and unbiased regions of
the extended 1D quantum waveguide. This reflection will be hereinafter
referred to as ``magnetic scattering'' associated with the above introduced
potential $V^{(reg)}(x)$. It can be easily verified that under constraint
\eqref{weak_magn} the reflection parameter \eqref{magn_refl} is modulo small
as compared to unity.

By substituting functions $\psi_{\pm}(x)$ in the form \eqref{psi_pm-in} into
\eqref{Green-Cochi}, the trial Green function whose both coordinate arguments
are located inside magnetically biased interval $\mathcal{L}$ is given as
\begin{equation}\label{Green-packets}
  G^{(V)}(x,x')=\mathcal{G}_1(x,x')\mathrm{e}^{i\varkappa(x-x')}+
          \mathcal{G}_2(x,x')\mathrm{e}^{-i\varkappa(x-x')}+
          \mathcal{G}_3(x,x')\mathrm{e}^{i\varkappa(x+x')}+
          \mathcal{G}_4(x,x')\mathrm{e}^{-i\varkappa(x+x')}\ ,
\end{equation}
where the last two terms in r.h.s. have emerged as a direct consequence of
the mode backscattering. In \eqref{Green-packets}, smooth envelope functions
are
\begin{subequations}\label{Green_smooth}
\begin{eqnarray}
  \label{Green_smooth-1}
 && \mathcal{G}_1(x,x')=\frac{-i}{2\varkappa}\mathcal{A}(x)
    \left[\Theta_+\frac{\pi_-(x')}{\pi_-(x)}-
    \Theta_-\frac{\gamma_+(x')}{\pi_+(x)}\Gamma_-(x)\mathrm{e}^{2i\varkappa L}\right]\ , \\
  \label{Green_smooth-2}
 && \mathcal{G}_2(x,x')=\frac{-i}{2\varkappa}\mathcal{A}(x)
    \left[\Theta_-\frac{\pi_+(x')}{\pi_+(x)}-
    \Theta_+\Gamma_+(x)\frac{\gamma_-(x')}{\pi_-(x)}\mathrm{e}^{2i\varkappa L}\right]\ , \\
  \label{Green_smooth-3}
 && \mathcal{G}_3(x,x')=\frac{-1}{2\varkappa}\mathcal{A}(x)
    \mathrm{e}^{i\varkappa L}\left[\Theta_+\frac{\gamma_-(x')}{\pi_-(x)}+
    \Theta_-\frac{\pi_+(x')}{\pi_+(x)}\Gamma_-(x)\right]\ , \\
  \label{Green_smooth-4}
 && \mathcal{G}_4(x,x')=\frac{-1}{2\varkappa}\mathcal{A}(x)
    \mathrm{e}^{i\varkappa L}\left[\Theta_-\frac{\gamma_+(x')}{\pi_+(x)}+
    \Theta_+\Gamma_+(x)\frac{\pi_-(x')}{\pi_-(x)}\right]\ ,
\end{eqnarray}
\end{subequations}
where the notations are used
\begin{subequations}\label{A_G}
\begin{align}\label{A_G-A}
  &\mathcal{A}(x)=\left[1+\Gamma_+(x)\Gamma_-(x)\mathrm{e}^{2i\varkappa
  L}\right]^{-1}\ , \\
  \label{A_G-Gamma}
  &\Gamma_{\pm}(x)=\gamma_{\pm}(x)/\pi_{\pm}(x)\ ,\\
  &\Theta_{\pm}=\theta[\pm(x-x')]\ .
\end{align}
\end{subequations}

The functions $\Gamma_{\pm}(x)$ play a particular role in the averaging
technique. It can be deduced from Eq.~\eqref{psi_pm-in} that these functions
are nothing but the reflection coefficients of spatial harmonics
$\pm\varkappa$ which are incident at the point $x$ onto the disordered layers
whose end coordinates are $x$ and $\pm L/2$, respectively. These reflection
factors are well-known to obey the Riccati-type dynamic equations
\cite{bib:Klyats86},
\begin{equation}\label{Riccati}
  \pm\frac{\mathrm{d}\Gamma_{\pm}(x)}{\mathrm{d}x}=
  2i\eta(x)\Gamma_{\pm}(x)-\zeta_{\pm}(x)+\zeta_{\pm}^*(x)\Gamma_{\pm}^2(x)\ ,
\end{equation}
which in our case must be supplied with ``initial'' conditions (one-side BC)
stemming from \eqref{bound_cond}, viz.
\begin{equation}\label{BC-Gamma_pm}
  \Gamma_{\pm}(\pm L/2)=\EuScript{R}^{(B)}\ .
\end{equation}

\subsection{Configuration averaging of the function \eqref{Green-packets}}

The averaging technique as applied to functionals of smoothed potentials
\eqref{eff_fields} was elaborated in
Refs.~\cite{bib:Tar00,bib:MakTar01,bib:FreiTar01}. It relies basically on the
fact that under WS conditions the functional arguments \eqref{eff_fields} can
be regarded as Gaussian-distributed random variables \cite{bib:LGP82}. Taking
account of correlation equalities \eqref{imp_corr}, the binary correlators of
these effective potentials, which would suffice to be allowed for, can be
cast to the form
\begin{subequations}\label{eff_fields-corr}
\begin{align}
  \label{eff_fields-corr-eta}
  & \big<\eta(x)\eta(x')\big>=\frac{1}{L_f}F_l(x-x')\ ,\\
  \label{eff_fields-corr-zeta}
  & \big<\zeta_{\pm}(x)\zeta_{\pm}^*(x')\big>=\frac{1}{L_b}F_l(x-x')\ ,
\end{align}
\end{subequations}
where $L_f$ and $L_b$ are the forward and backward scattering lengths given
in \eqref{ext_length} for the particular mode $\bm{\nu}$. The function
\begin{equation}\label{F_l(x)}
  F_l(x)=\int_{-\infty}^{\infty}\frac{\mathrm{d}q}{2\pi}\mathrm{e}^{iqx}
  \frac{\sin^2(ql)}{(ql)^2}=\frac{1}{2l}\left(1-\frac{|x|}{2l}\right)
  \theta(2l-|x|)
\end{equation}
is sharp on a scale of significant change in smooth amplitudes of electron
wave functions, and therefore it can act as the under-limiting
$\delta$-function when averaging amplitude factors $\mathcal{G}_i(x,x')$
in~\eqref{Green-packets}. Prior to averaging these factors, it is
advantageous to absorb, wherever possible, the forward-scattering field
$\eta(x)$ by going over from functions $\gamma_{\pm}(x)$, $\pi_{\pm}(x)$ and
$\zeta_{\pm}(x)$ to phase-renormalized functions
\begin{subequations}\label{phase_renorm}
\begin{align}
  \label{phase_renorm-gamma}
 \widetilde{\gamma}_{\pm}(x)&= \gamma_{\pm}(x)
 \exp\left[\pm i\int_x^{\pm L/2}\mathrm{d}x_1\eta(x_1)\right]\ , \\
  \label{phase_renorm-pi}
 \widetilde{\pi}_{\pm}(x)&= \pi_{\pm}(x)
 \exp\left[\mp i\int_x^{\pm L/2}\mathrm{d}x_1\eta(x_1)\right]\ , \\
 \label{phase_renorm-zeta}
 \widetilde{\zeta}_{\pm}(x)&= \zeta_{\pm}(x)
 \exp\left[\pm 2i\int_x^{\pm L/2}\mathrm{d}x_1\eta(x_1)\right]\ .
\end{align}
\end{subequations}
An important thing is that correlation relation \eqref{eff_fields-corr-zeta}
remains unchanged after renormalization \eqref{phase_renorm-zeta}.

In view of short-range correlation of random functions \eqref{eff_fields} and
due to the causal nature of functionals being averaged, the averaging of
functionals with different sign indices in \eqref{Green_smooth} can be done
separately. By applying the Furutsu-Novikov formula for Gaussian random
processes \cite{bib:Klyats86}, we first average the equation for renormalized
reflection factors
$\widetilde{\Gamma}_{\pm}(x)=\widetilde{\gamma}_{\pm}(x)/\widetilde{\pi}_{\pm}(x)$,
i.~e.
\begin{equation}\label{renormGamma}
  \pm\frac{\mathrm{d}\widetilde{\Gamma}_{\pm}(x)}{\mathrm{d}x}=
  -\widetilde{\zeta}_{\pm}(x)+
  \widetilde{\zeta}_{\pm}^*(x)\widetilde{\Gamma}_{\pm}^2(x)\ .
\end{equation}
With condition \eqref{BC-Gamma_pm} taken into account this yields
\begin{equation}\label{AvGamma_pm-sol}
  \langle\widetilde{\Gamma}_{\pm}(x)\rangle=\EuScript{R}^{(B)}
  \exp\left[-\frac{1}{L_b}\left(\frac{L}{2}\mp x\right)\right]\ .
\end{equation}
In view of WMS-motivated smallness of the reflection parameter
$\EuScript{R}^{(B)}$, this implies that only the terms that do not contain
the factors of $\widetilde{\Gamma}_{\pm}(x)$ and
$\widetilde{\gamma}_{\pm}(x)$ can be kept in \eqref{Green_smooth} when
averaging Green function \eqref{Green-packets}.

Next, in order to average the ratio
$\widetilde{\pi}_{\pm}(x')/\widetilde{\pi}_{\pm}(x)$ consider its Fourier
transform over $x'$ coordinate,
\begin{equation}\label{Fourier-1}
  \Phi^{(\pm)}(x,q)=\pm\int_x^{\pm L/2}\mathrm{d}x_1
  \frac{\widetilde{\pi}_{\pm}(x_1)}{\widetilde{\pi}_{\pm}(x)}
  \exp\left[-iq(x-x_1)+i\varkappa|x-x_1|
  \pm i\int_{x_1}^x\mathrm{d}x_2\eta(x_2)\right]\ ,
\end{equation}
where we have separated the forward-scattering random field $\eta(x)$ in the
form of a phase factor. Averaging over this field yields readily
\begin{equation}\label{Av_eta}
  \Big<\exp\left[\pm i\int_{x_1}^x\mathrm{d}x_2\eta(x_2)\right]\Big>_{\eta}=
  \exp\left(-\frac{|x-x_1|}{2L_f}\right)\ .
\end{equation}
Then, after averaging over $\eta(x)$, the function \eqref{Fourier-1} comes to
obey the equation
\begin{equation}\label{Phi-eq}
  \mp\frac{\mathrm{d}\big<\widetilde{\Phi}^{(\pm)}(x,q)\big>_{\eta}}{\mathrm{d}x}=
  1-\left(\frac{1}{2L_f}-i\varkappa\mp iq\right)\big<\widetilde{\Phi}^{(\pm)}(x,q)\big>_{\eta}-
  \widetilde{\zeta}_{\pm}^*(x)\widetilde{\Gamma}_{\pm}(x)\big<\widetilde{\Phi}^{(\pm)}(x,q)\big>_{\eta}
  \ ,
\end{equation}
which is to be solved jointly with Eq.~\eqref{renormGamma}. Averaging of
Eq.~\eqref{Phi-eq} over the effective back-scattering field
$\widetilde{\zeta}_{\pm}(x)$ results in the dynamic equation
\begin{equation}\label{Phi_aver-eq}
  \mp\frac{\mathrm{d}\big<\widetilde{\Phi}^{(\pm)}(x,q)\big>}{\mathrm{d}x}=
  1-\left[\frac{1}{2}\left(\frac{1}{L_f}+\frac{1}{L_b}\right)-
  i\varkappa\mp iq\right]\big<\widetilde{\Phi}^{(\pm)}(x,q)\big>\ ,
\end{equation}
which is to be solved with obvious ``initial'' condition,
$\big<\widetilde{\Phi}^{(\pm)}(\pm L/2,q)\big>=0$. The solution to
Eq.~\eqref{Phi_aver-eq} yields
\begin{equation}\label{Phi-sol}
  \big<\widetilde{\Phi}^{(\pm)}(x,q)\big>=
  \left[\frac{1}{2}\left(\frac{1}{L_f}+\frac{1}{L_b}\right)-
  i\varkappa\mp iq\right]^{-1}
  \Bigg[1-\exp\left\{-\left[\frac{1}{2}\left(\frac{1}{L_f}+\frac{1}{L_b}\right)-
  i\varkappa\mp iq\right]\left(L/2\mp x\right)\right\}\Bigg]\ ,
\end{equation}
wherefrom the result is obtained
\begin{equation}\label{Green_1-2terms}
  \big<\mathcal{G}_1(x,x')\big>\mathrm{e}^{i\varkappa(x-x')}+
  \big<\mathcal{G}_2(x,x')\big>\mathrm{e}^{-i\varkappa(x-x')}=
  \frac{-i}{2\varkappa}
  \exp\Bigg[\mathrm{i}\varkappa|x-x'|-\frac{1}{2}
  \left(\frac{1}{L_f}+
  \frac{1}{L_b}\right)|x-x'|\Bigg] \ .
\end{equation}

The terms \eqref{Green_smooth-3} and \eqref{Green_smooth-4} can be averaged
in the same manner. The result is much more cumbersome, namely
\begin{subequations}\label{G3_G4}
\begin{align}
 \label{G3}
 \big<\mathcal{G}_3(x,x')\big>=-\frac{\EuScript{R}^{(B)}}{2\varkappa}
 \exp\left[i\varkappa L-L\left(\frac{1}{L_f}+\frac{1}{2L_b}\right)\right]
 \Bigg\{&\theta(x-x')\exp\left[-\frac{x}{2}\left(\frac{1}{L_f}+
 \frac{1}{L_b}\right)-\frac{x'}{2}\left(\frac{3}{L_f}+\frac{1}{L_b}\right)\right]
 \notag\\
 +\,&\theta(x'-x)\exp\left[-\frac{x'}{2}\left(\frac{1}{L_f}+
 \frac{1}{L_b}\right)-\frac{x}{2}\left(\frac{3}{L_f}+\frac{1}{L_b}\right)\right]
 \Bigg\}
\end{align}
\begin{align}
 \label{G4}
 \big<\mathcal{G}_4(x,x')\big>=-\frac{\EuScript{R}^{(B)}}{2\varkappa}
 \exp\left[i\varkappa L-L\left(\frac{1}{L_f}+\frac{1}{2L_b}\right)\right]
 \Bigg\{&\theta(x-x')\exp\left[\frac{x'}{2}\left(\frac{1}{L_f}+
 \frac{1}{L_b}\right)+\frac{x}{2}\left(\frac{3}{L_f}+\frac{1}{L_b}\right)\right]
 \notag\\
 +\,&\theta(x'-x)\exp\left[\frac{x}{2}\left(\frac{1}{L_f}+
 \frac{1}{L_b}\right)+\frac{x'}{2}\left(\frac{3}{L_f}+\frac{1}{L_b}\right)\right]
 \Bigg\}\ .
\end{align}
\end{subequations}
Fortunately, in the WMS limit, because of the factor $\EuScript{R}^{(B)}$
being small, these terms may be omitted, leading finally to result
\eqref{G^V_nu}.


\end{document}